\documentclass[a4paper,12pt]{article}
\usepackage[papersize={216mm,330mm},tmargin=15mm,bmargin=15mm,lmargin=15mm,rmargin=15mm]{geometry}
\usepackage{fancyhdr}
\usepackage{amsmath}
\usepackage{amssymb}
\usepackage{graphicx,multicol}
\usepackage{subfigure}
\usepackage{bm}
\usepackage{tabu}
\usepackage[table]{xcolor}
\usepackage{hyperref}
\usepackage{textgreek}
\usepackage{upgreek}
\usepackage{multicol}
\usepackage{natbib}
\usepackage{comment}
\usepackage{setspace}
\usepackage{siunitx}
\usepackage[utf8]{inputenc}
\usepackage[T1]{fontenc}
\setcitestyle{authoryear}
\providecommand{\keywords}[1]
{
  \small
  \textbf{\textit{Keywords---}} #1
}

\title{Gamma-Band Correlations in Primary Visual Cortex}

\usepackage{authblk}
\author[1,2]{X. Liu \thanks{Corresponding Author: xliu9362@uni.sydney.edu.au (X.Liu)}}
\author[1,2]{P. Sanz-Leon}
\author[1,2]{P. A. Robinson}

\affil[1]{School of Physics, University of Sydney, New South Wales 2006, Australia}
\affil[2]{Center for Integrative Brain Function, University of Sydney, New South Wales 2006, Australia}

\begin{document}
\maketitle

\begin{abstract}
Neural field theory is used to quantitatively analyze the two-dimensional spatiotemporal correlation properties of gamma-band (30 -- 70 Hz) oscillations evoked by stimuli arriving at the primary visual cortex (V1), and modulated by patchy connectivities that depend on orientation preference (OP).
Correlation functions are derived analytically under different stimulus and measurement conditions. The predictions reproduce a range of published experimental results, including the existence of two-point oscillatory temporal cross-correlations with zero time-lag between neurons with similar OP, the influence of spatial separation of neurons on the strength of the correlations, and the effects of differing stimulus orientations.
\end{abstract}
\maketitle
\doublespacing
\thispagestyle{empty}

\section{Introduction}
The primary visual cortex (V1) is the first cortical area to process visual inputs that arrive from the retina  via the lateral geniculate nucleus of the thalamus (LGN), and it feeds the processed signals forward to higher visual areas, and back to the LGN. The feed-forward visual pathway from the eyes to V1 is such that the neighboring cells in V1 respond to neighboring regions of the retina \citep{schiller_visual}.
V1 can be approximated as a two-dimensional layered sheet \citep{Tovee_visual_system}. Neurons that span vertically through multiple layers of V1 form a functional cortical column, and these neurons respond most strongly to a preferred stimulus orientation, right or left eye, direction of motion, and other feature preferences. Thus, various features of the visual inputs are mapped to V1 in different ways. These maps are overlaid such that a single neural cell responds to several features and all preferences within a given visual field are mapped to a small region of V1, often termed a hypercolumn, which corresponds to a particular visual field in the overall field of vision \citep{Hubel_column_arrangm_1962,hubel_sequence_1974,Miikkulainen_visual_maps}.

\begin{figure}[h!]
\centering
\includegraphics[width=0.90 \textwidth]{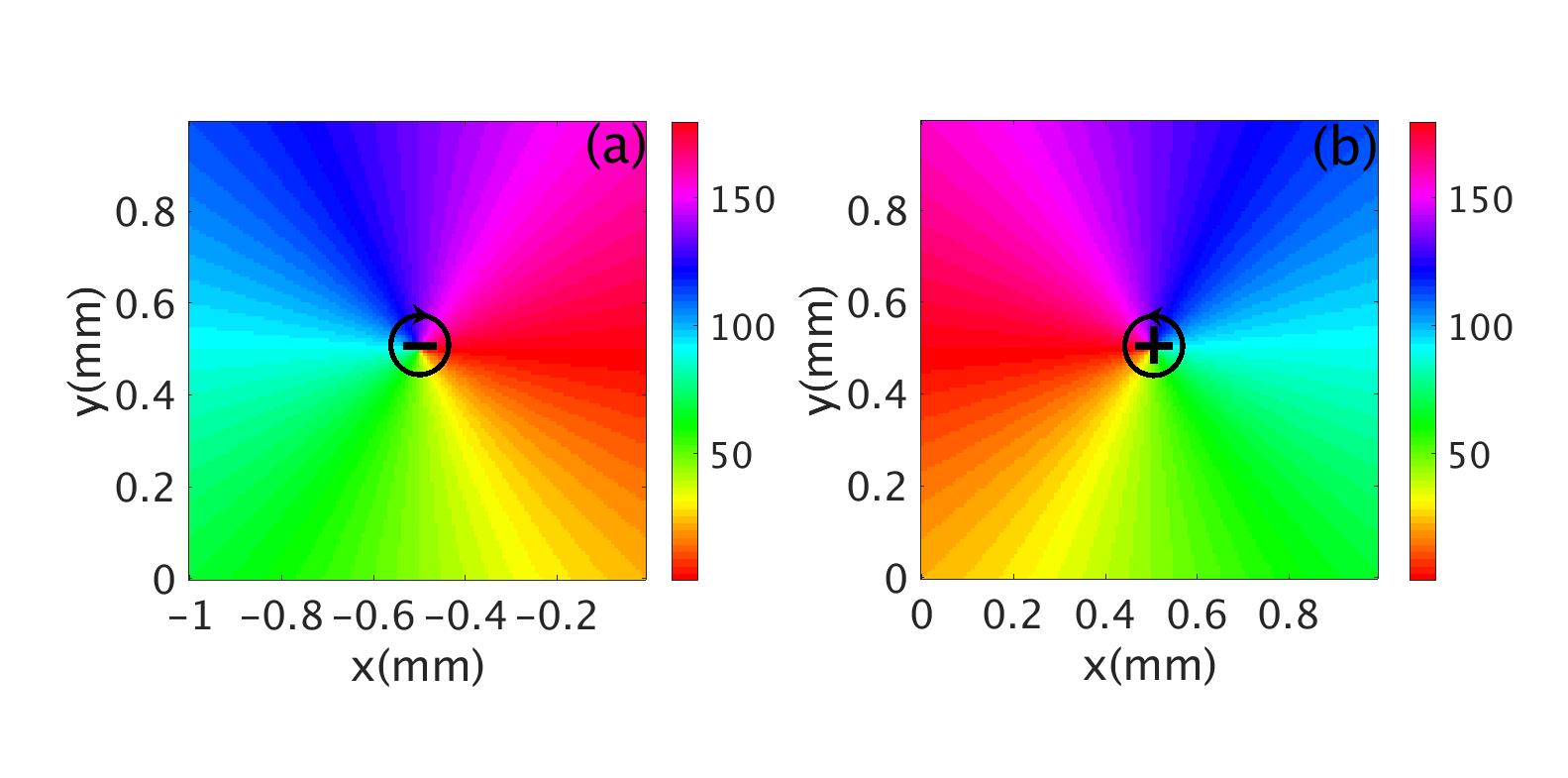}
\includegraphics[width=0.4\textwidth]{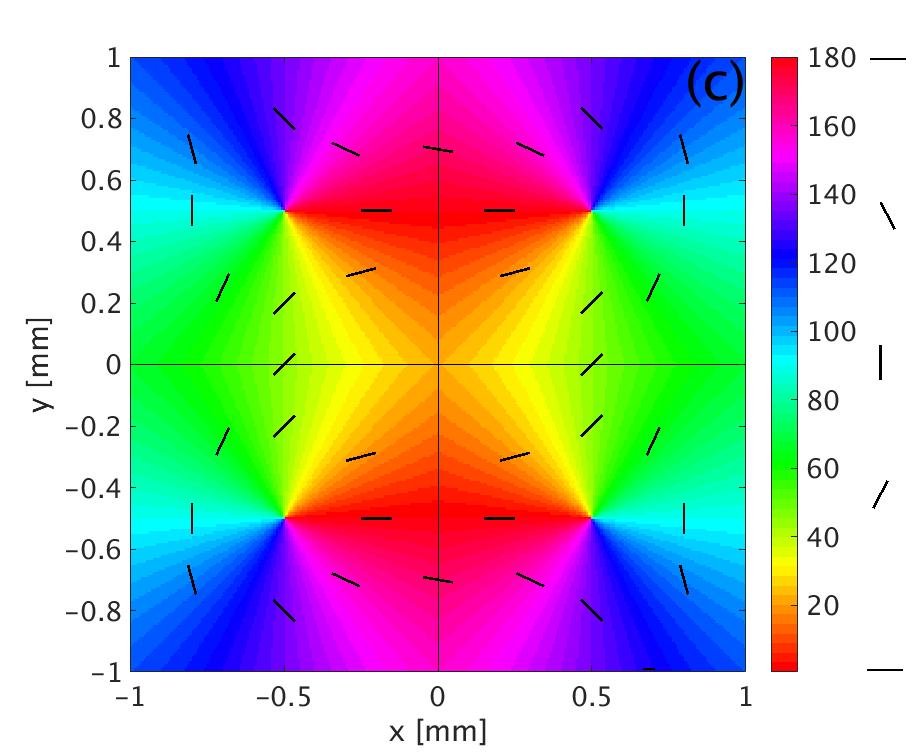}%
\includegraphics[width=0.4\textwidth]{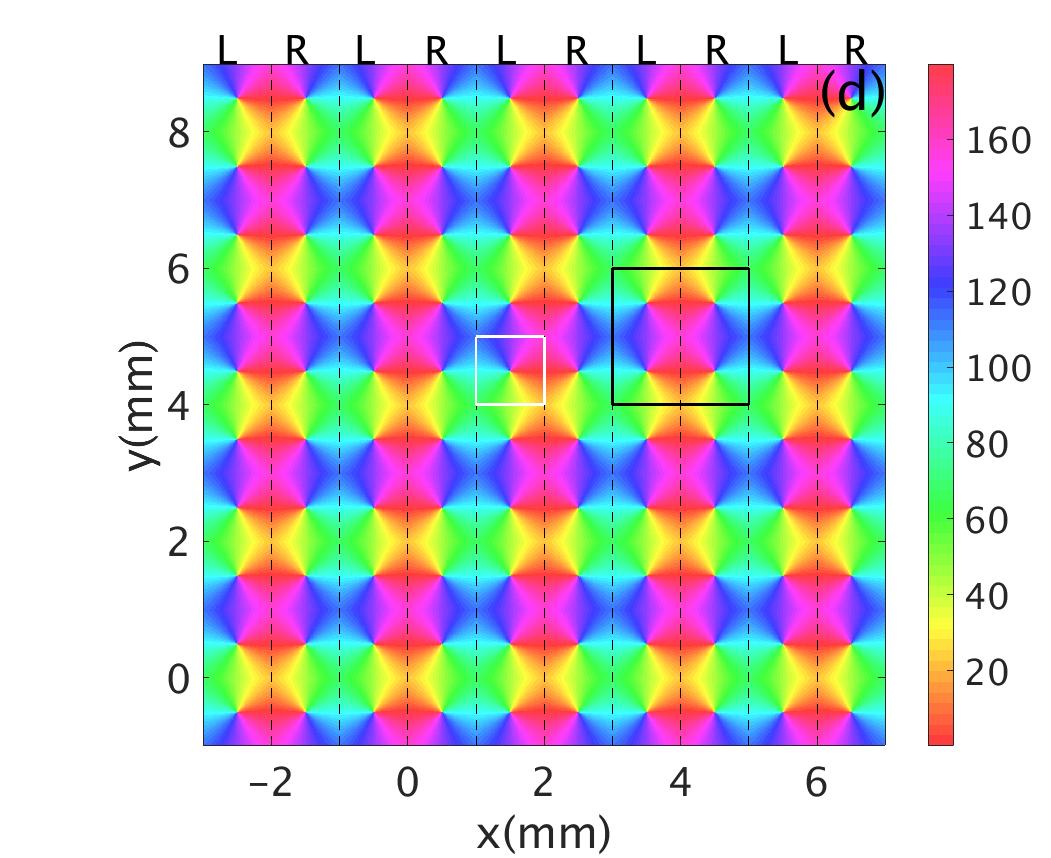}
\caption{Schematics of visual feature preference maps in V1 with color bars indicating OP in degrees. (a) Negative pinwheel. (b) Positive pinwheel. (c) Lattice unit cell (hypercolumn). The vertical line divides the unit cell into left and right OD columns of equal width, while the horizontal and vertical lines split the unit cell into four squares, each containing one OP pinwheel. The short bars highlight the OP at various locations. (d) Periodic spatial structure of OP and OD columns across a small piece of V1 comprising 25 unit cells. Dashed lines bound left (L) and right (R) OD columns. One pinwheel is outlined in white and one unit cell is outlined in black. Frames (a) and (b) are adapted from \cite{Kang_pinwheels}.}
 \label{fig:op_map}
\end{figure}

A prominent feature of V1 is the presence of ocular dominance (OD) stripes, which reflect the fact that left- and right-eye inputs are mapped to alternating stripes $\sim 1$ mm wide, with each hypercolumn including left- and right-eye OD regions. Orientation preference (OP) of neurons for particular edge orientations in a visual field is mapped to regions within each hypercolumn such that neurons with particular OP are located adjacent to one another and OP spans the range from $0^\circ$ to $180^\circ$. Typically, OP varies with azimuth relative to a center, or singularity, in the hypercolumn in an arrangement called a pinwheel. The OP angle in each pinwheel rotates either clockwise (negative pinwheel) or counterclockwise (positive pinwheel), and neighboring pinwheels have opposite signs \citep{blasdel_orientation_1992,Braitenberg1979,gotz_d-blob_1987,Gotz_1988,swindale_review_1996}. Hence, a hypercolumn must have left and right OD stripes with positive and negative pinwheels in each, as suggested by \cite{bressloff_crystal} and \cite{Veltz_2015}.
In Figs~\ref{fig:op_map}(a) and (b) we illustrate a negative pinwheel and a positive pinwheel, respectively, while Figs~\ref{fig:op_map}(c) and (d) show a hypercolumn containing four pinwheels, and an array of such hypercolumns, respectively. In such an array, the hypercolumn is the unit cell of the lattice and the schematic resembles maps reconstructed from \textit{in-vivo} experiments, although the stripes have been approximated as straight here  \citep{blasdel_orientation_1992,bonhoeffer_iso-orientation_1991,bonhoeffer_layout_1993,obermayer_geometry_1993}.

An additional feature of V1 is that regions of similar OP are preferentially linked within and between unit cells by patchy lateral connections \citep{Gilbert_patchy_connectn,Rockland_Lund_1982}. Furthermore, patchy connections into and out of a given OP region
are concentrated along an axis that points in the direction of the OP.  This means that cells that are sensitive to a contour of given orientation preferentially project to (and receive projections from) cells of similar OP located along the continuation of that contour, which has been argued to be important to the completion of occluded contours and the binding problem \citep{Miikkulainen_visual_maps,stemmler_lateral,LOFFLER_perception,Li_integration}. Most notably, the projections from a given unit cell depend strongly on the OP at the source neurons within that cell and are thus strongly anisotropic  \citep{bosking_orientation_1997}.

When one considers activity in V1, numerous experiments and studies \citep{eckhorn_coherent_1988,konig_relation_1995,singer_visual_1995,engel_inter-columnar_1990,hata_horizontal_interaction_1991,gray_oscillatory_response_1989} have shown that neurons with similar feature preference in V1 exhibit synchronized gamma band (30 -- 70 Hz) oscillations when the stimulus is optimal, by measuring the multi-unit activities (MUA) and local field potentials (LFP) in area 17 of cats using multi-electrodes. They also showed that the corresponding two-point correlation functions of MUA or LFP commonly have peaks at zero time-lag. Moreover, these synchronized gamma oscillation in V1 arise from the spatial structure of V1, modulated by the specific feature preferences involved. It also has been argued that such synchronized oscillation in gamma band may be involved in visual perception, the binding of related features into unified percepts, and the occurrence of visual hallucinations \citep{Gary_stimulus_1990,engel_dynamic_2001,bressloff_what_2002,siegel_cortical_2011,Henke_hallucination}.

Previous theoretical studies \citep{robinson_propagator_2005,robinson_patchy_2006,robinson_visual_2007} used neural field theory (NFT) with patchy propagators to show that patchy connectivity could support gamma oscillations with correlation properties whose features resembled those of some of the experiments noted above. However, the effect of OP on the patchy propagators was not incorporated and the correlations were only explored as functions of one spatial dimension.

In this paper, we generalize and explore the spatiotemporal correlation functions of \cite{robinson_patchy_2006,robinson_visual_2007} to two spatial dimensions, and account for the effect of OP on the patchy propagators. We then compare the resulting spatiotemporal correlations with several MUA experiments.
In Sec.~\ref{sec:theory}, we briefly describe the relevant aspects of NFT including patchy propagators. In Sec.~\ref{sec:correlations_theory}, we derive the general 2D correlation function in V1 via the linear NFT transfer function of V1. Section~\ref{sec:propagation} describes a spatial propagator, which modulates the connection strength between cortical locations that have similar feature preference, and the Fourier coefficients of this propagator are applied to the numerical calculation of the correlation properties. The properties of these correlation functions are explored in Sec.~\ref{sec:temporal_corr_properties}, including their predictions for oscillation frequency, time decay, effects of the spatial separation between the measurement points, and the modulation by the OP in V1. The predictions compared with specific experimental outcomes in Sec.~\ref{sec:compare_results}, and the results are summarized and discussed in Sec.~\ref{sec:discussion}.

\section{Theory}
\label{sec:theory}
In order to analyze correlations in the patchily connected cortex, we first briefly review an established neural field model of the relevant corticothalamic system in Sec.~\ref{subsec:nft}, and calculation of its approximate transfer function in the gamma frequency range of several tens of Hz, with further details of the derivations available in prior papers \citep{robinson_patchy_2006,robinson_visual_2007}. In Secs~\ref{sec:correlations_theory} and \ref{sec:propagation} we generalize the patchy connectivity to two dimensions (2D) and calculate the resulting 2D correlation functions in order to treat the effects of both OD and OP together.

\subsection{Neural Field Theory}
\label{subsec:nft}
\begin{figure}[h]
    \centering
    \includegraphics[width=0.5\textwidth]{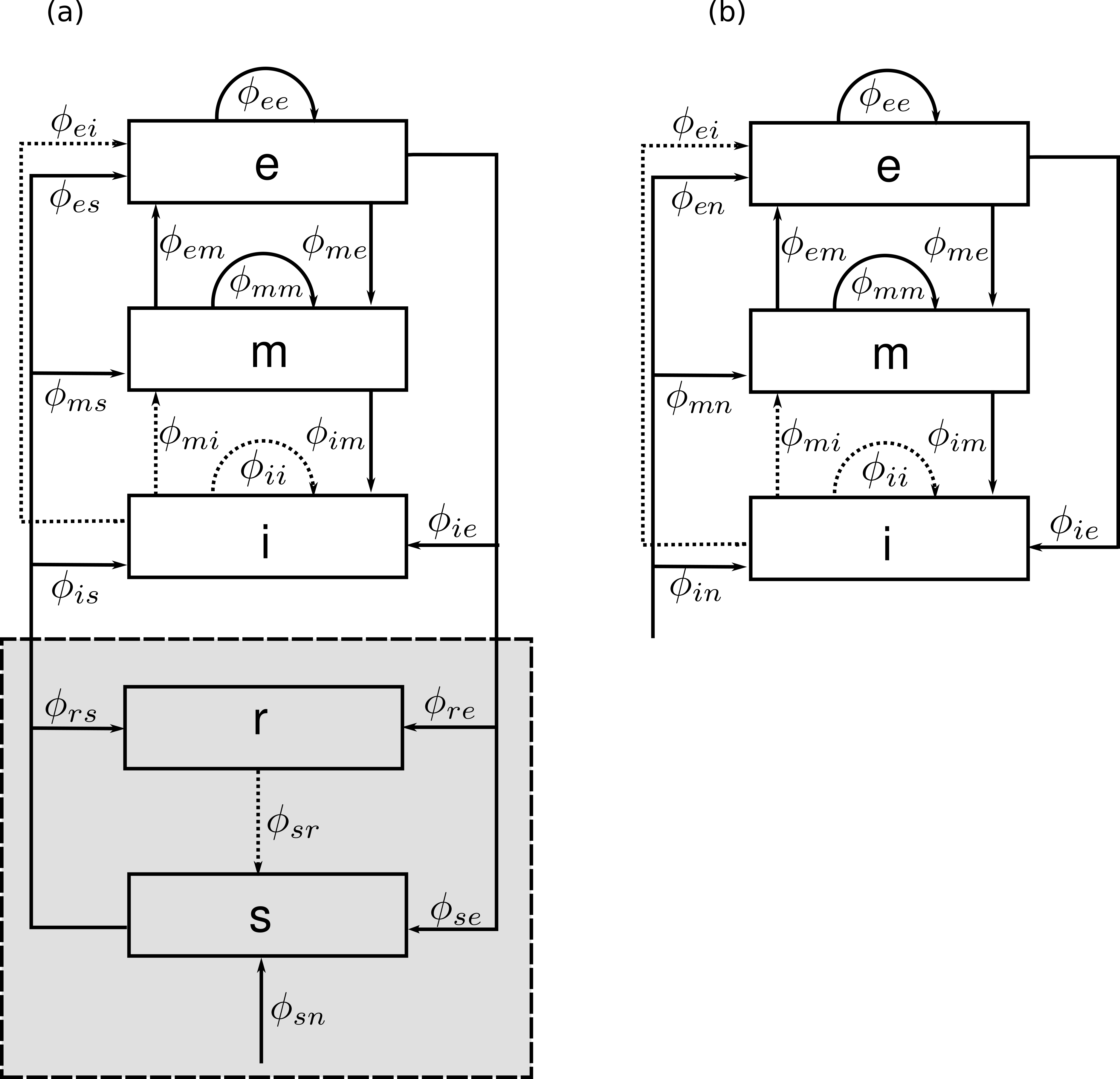}
    \caption{Schematics of the corticothalamic system. (a) The full EMIRS model with the thalamus shown in the gray rectangle, each $\phi_{ab}$ quantifies the connection to population a from population b. (b) The simplified EMIRS model with the thalamic part approximated as a cortical input.}
    \label{fig:emirs}
\end{figure}

The previously developed corticothalamic model \citep{robinson_propagator_2005} treats five neural populations, which are the long-range excitatory pyramidal neurons (\textit{e}), midrange patchy excitatory neurons (\textit{m}), short-range inhibitory interneurons (\textit{i}), thalamic reticular neurons (\textit{r}), and thalamic relay neurons (\textit{s}); hence, it is termed the EMIRS model. Figure \ref{fig:emirs}(a) shows the full EMIRS model and its connectivities between neural populations, including the axonal fields (described further below) $\phi_{ab}$ of spike rates arriving at neurons of population $a$ from those of population $b$, where $a,b=e,m,i,r,s,n$. The external input signal $\phi_{sn}$ is incident on the relay nuclei.

In this work, we are mainly concerned with cortical neural activities in the gamma band (\SI{30}{Hz} -- \SI{70}{Hz}), which are  higher than the resonant frequency ($\sim$\SI{10}{Hz}) of the corticothalamic loops. This enables us to neglect the corticothalamic feedback loops of the full EMIRS model, leading to the reduced model in Fig.~\ref{fig:emirs}(b). This model only includes the cortical excitatory, mid-range, and short-range inhibitory populations, and the signals from the thalamus are treated as the input to the cortex. Thus, rather than having feedback inputs from the thalamus, we approximate these inputs as a common external input $\phi_{an}$ to the cortex. The subscript $a$ denotes the three cortical neural populations (\textit{e}, \textit{m}, \textit{i}).

Normal brain activity has been widely modeled as corresponding approximately to linear perturbations from a fixed point, with successful applications to experiments such as electroencephalographic (EEG) spectra, evoked response potentials, visual hallucinations, and other phenomena \citep{Henke_hallucination,Robinson_1998,Robinson_2002,Robinson_EEG_2004}. Hence, in the present work, we restrict attention to the linear regime, which is justified so long as stimuli are not too strong.

NFT averages neural properties and activity over a linear scale of a few tenths of a millimeter to treat the dynamics on larger scales, which is appropriate for the present applications \citep{Deco_2008,Robinson_multiscale_2005}.

Cells with voltage-gated ion channels produce action potentials when the soma potential exceeds a threshold $\theta_{a}$. In the linear regime, changes $Q_{a}$ in the mean population firing rate are related to the mean soma potential $V_{a}$ by
\begin{equation}
Q_{a}(\mathbf{k},\omega)=\rho_{a}V_{a}(\mathbf{k},\omega),
\end{equation}
where $\rho_a$ is a constant.

The mean linear perturbation $V_{a}$ to the soma potential of neurons $a$ is approximated by summing contributions $V_{ab}$ that resulting from activities of all types of synapse on neurons in the spatially extended population $a$ from those of type $b$. Thus,
\begin{equation}\label{eq_v_a}
V_{a}(\mathbf{r},t)=\sum_{b}V_{ab}(\mathbf{r},t),
\end{equation}
where $\mathbf{r}$ is the spatial location on the cortex, approximated as a 2D sheet, and $t$ is the time. In the Fourier domain, Eq.~\eqref{eq_v_a} can be written as
\begin{equation}
V_{a}(\mathbf{k},\omega)=\sum_{b}V_{ab}(\mathbf{k},\omega),
\end{equation}
where we define the Fourier transform and its inverse via
\begin{equation}
g(\mathbf{k},\omega)=\int d^{2}\mathbf{r} \int dt\, g(\mathbf{r},t)e^{i\omega t - i\mathbf{k \cdot r}},
\end{equation}
\begin{equation}
g(\mathbf{r},t)=\int \frac{d^{2}\mathbf{k}}{(2\pi )^{2}} \int \frac{d \omega}{(2\pi)}\, g(\mathbf{k},\omega)e^{i\mathbf{k \cdot r}-i\omega t}.
\end{equation}

Due to the dependence of $V_{ab}$ on the synaptic dynamics, signal dispersion in the dendrites, and soma charging, the soma potential corresponding to a delta function input can be approximated by
\begin{equation}\label{eq_Vab}
V_{ab}(\mathbf{k},\omega)=L_{ab}(\omega)P_{ab}(\mathbf{k},\omega),
\end{equation}
where $P_{ab}$ is the arrival rate of incoming spikes, and $L_{ab}$ is the synapse-to-soma transfer function, with
\begin{equation}\label{L_ab}
L_{ab}(\omega)={}\left (1-i\omega/\alpha_{ab}\right)^{-1}\left (1-i\omega/\beta_{ab}\right)^{-1},
\end{equation}
where $\alpha_{ab}$ and $\beta_{ab}$ are the decay and rise rates of the soma response, respectively.

In Eq.~\eqref{eq_Vab}, $P_{ab}$ depends on $Q_{b}$ at various source locations and earlier times \citep{robinson_visual_2007}, whose influences $\phi_{ab}$ propagate to $a$ from $b$ via axons, with
\begin{equation}\label{eq_P_ab}
P_{ab}(\mathbf{k},\omega)=\hat\nu_{ab}(\mathbf{k},\omega)\phi_{ab}(\mathbf{k},\omega),
\end{equation}
\begin{equation}
\label{phi_ab}
\phi_{ab}(\mathbf{k},\omega)=e^{i\omega\tau_{ab}}\Gamma_{ab}(\mathbf{k},\omega)Q_{b}(\mathbf{k},\omega),
\end{equation}
where $\Gamma_{ab}$ describes axonal propagation.
In Eq.~\eqref{phi_ab}, $\tau_{ab}$ is the time delay between spatially discrete neuron populations (i.e., not between different {\bf r} on the cortex) and $\hat\nu_{ab}$ represents the coupling of $\phi_{ab}$ to population $a$. In the simplest case of proportional coupling,
\begin{equation}
\hat\nu_{ab}(\mathbf{k},\omega)=N_{ab}s_{ab},
\end{equation}
where $N_{ab}$ is the mean number of synaptic connections to each neuron of type $a$ from neurons of type $b$ and $s_{ab}$ is their mean strength. More generally, $\hat\nu$ can describe couplings that are sensitive to other features of $\phi_{ab}$, such as spatial or temporal derivatives, which can increase sensitivity to features such as edges in the visual stimulus \citep{robinson_propagator_2005,robinson_patchy_2006,robinson_visual_2007}.

Axonal propagation can be approximately described by a damped wave equation \citep{jirsa_field_1996,Robinson_1997,schiff_dynamical_2007}
\begin{equation} \label{eq_damp_wave}
\left [ \frac{1}{\gamma_{ab}^2}\frac{\partial^2 }{\partial t^2}+\frac{2}{\gamma_{ab}}\frac{\partial }{\partial t}+1-r_{ab}^2 \triangledown ^2\right ]\phi_{ab}(\mathbf{r},t)=Q_{b}(\mathbf{r},t),
\end{equation}
where $\gamma_{ab}=v_{ab}/r_{ab}$ is the temporal damping coefficient, $v_{ab}$ is the wave velocity, and $r_{ab}$ is the characteristic range of axons that project to population $a$ from $b$. In Fourier space, in the absence of patchy connections, one has \citep{robinson_propagator_2005}
\begin{equation}\label{eq_prop_uniform}
\Gamma_{ab}^{(0)}(\mathbf{k}, \omega)={}\frac{1}{(k^{2}+q_{0ab}^{2})r_{ab}^{2}},
\end{equation}
\begin{equation}\label{eq_prop_uni_2}
q_{0ab}^{2}r_{ab}^{2}={}(1-i\omega/\gamma_{ab})^{2}.
\end{equation}

To incorporate the patchy propagation, we approximate the OP-OD structure of V1 as being periodic, which results in periodic spatial modulation of the propagator in Eq.~\eqref{phi_ab}, giving \citep{robinson_visual_2007}
\begin{equation}\label{eq_general_prop}
\Gamma_{ab}(\mathbf{k}, \omega)={}\sum_{\mathbf{K}}c_{\mathbf{K}}\Gamma_{ab}^{(0)}(\mathbf{k-K}, \omega),
\end{equation}
where the $c_{\mathbf{K}}$ are the Fourier coefficients of the function that describes the spatial feature preference (i.e., OP and/or OD), and $\mathbf{K}$ ranges over the reciprocal lattice vectors of the periodic structure \citep{robinson_visual_2007}. We analyze the $c_{\bf K}$ in Sec.~\ref{sec:propagation} below.

In order to perform further linear analysis of the system, we write $Q_{a}(\mathbf{k},\omega)$ and $Q_{b}(\mathbf{k},\omega)$ via Eqs \eqref{eq_v_a} -- \eqref{eq_P_ab}, which yields the set of linear equations
\begin{equation}\label{eq_Q_a}
Q_{a}(\mathbf{k},\omega)={}\sum_{b}X_{ab}(\mathbf{k},\omega)Q_{b}(\mathbf{k},\omega),
\end{equation}
with
\begin{equation}\label{eq_h_ab}
X_{ab}(\mathbf{K},\omega)= J_{ab}(\mathbf{k},\omega)\Gamma_{ab}(\mathbf{k},\omega),
\end{equation}
\begin{equation}\label{eq_J_ab}
J_{ab}(\mathbf{k},\omega)=\rho_{a}L_{ab}(\omega)\nu_{ab}(\mathbf{k},\omega)e^{i\omega\tau_{ab}}.
\end{equation}

\subsection{System Transfer Function and Resonances}\label{linear_analysis_gamma_osci}
Turning to the system in Fig.~1, Eqs~\eqref{eq_Q_a} -- \eqref{eq_J_ab} can be used to write the activity changes $Q_e$ in the pyramidal neurons in terms of changes in the firing rate $Q_n$ that implicitly drives the input signal $\phi_{sn}$.
At gamma frequencies, where corticothalamic feedback is too slow to respond effectively, this was found to yield \citep{robinson_patchy_2006,robinson_visual_2007}
\begin{equation}\label{eq_T_en}
T_{en}(\mathbf{k},\omega)=
\frac{Q_{e}(\mathbf{k},\omega)}{Q_{n}(\mathbf{k},\omega)}={}\frac{X_{en}}{1-X_{ee}-X_{em}-X_{ei}}.
\end{equation}

Resonances of the system that determine spatiotemporal properties of the gamma oscillations arise from the poles of the transfer function, which correspond to zeros of the denominator of Eq.~(\ref{eq_T_en}). At millimeter scales, $k\gg1/r_{ee}$ and $|X_{ee}|\ll |X_{ei}|$, so the resonance condition becomes
\begin{equation}\label{eq_X_em}
1-X_{em}-X_{ei}=0.
\end{equation}
Substituting Eqs \eqref{L_ab}, \eqref{eq_prop_uniform}, \eqref{eq_h_ab}, and \eqref{eq_J_ab} into  Eq. \eqref{eq_X_em} gives
\begin{equation}\label{eq_simplify_X_em}
\sum_{\mathbf{K}}\frac{\hat{G}(\mathbf{k},\omega)}{(\mathbf{k-K})^{2}r_{em}^{2}+(1-i\omega/\gamma_{em})^{2}}={}\left(1-\frac{i\omega}{\alpha_{em}}\right)\left(1-\frac{i\omega}{\beta_{em}}\right)- \frac{G_{ei}}{k^{2}r_{ei}^2+1} \,,
\end{equation}
where
\begin{equation}
\hat{G}(\mathbf{k},\omega)= c_{\mathbf{K}}\rho_{e}\hat{\nu}_{em}(\mathbf{k},\omega).
\end{equation}
When $\mathbf{k}\approx \mathbf{K}$, the denominator on the left hand side of Eq.~\eqref{eq_simplify_X_em} is small, and the corresponding term dominates the sum over the lattice vectors $\mathbf{K}$. Assuming $\hat{G}(\mathbf{k},\omega)$ is purely spatial, $\hat{G}(\mathbf{k},\omega)$ can be written as $\hat{G}(\mathbf{K})$, so Eq.~\eqref{eq_simplify_X_em} becomes \citep{robinson_patchy_2006,robinson_visual_2007}
\begin{equation}
\frac{\hat{G}(\mathbf{K})}{(\mathbf{k-K})^{2}r_{em}^{2}+(1-i\omega/\gamma_{em})^{2}}={}\left(1-\frac{i\omega}{\alpha_{em}}\right)\left(1-\frac{i\omega}{\beta_{em}}\right)- \frac{G_{ei}}{K^{2}r_{ei}^2+1},
\end{equation}
\begin{equation}\label{eq_simplify_K}
\begin{aligned}
\hat{G}(\mathbf{K})={}&\left [\left(1-\frac{i\omega}{\alpha_{em}}\right)\left(1-\frac{i\omega}{\beta_{em}}\right)- \frac{G_{ei}}{K^{2}r_{ei}^2+1} \right ]\\
& \times \left [(\mathbf{k-K})^{2}r_{em}^{2}+\left(1-\frac{i\omega}{\gamma_{em}^{2}}\right)\right].
\end{aligned}
\end{equation}
\cite{robinson_visual_2007} showed that each value of ${\bf K}$ can yield a resonance with frequency
\begin{equation}\label{eqn_omega}
\Omega^2=\frac{\gamma\left[2\alpha\beta\left(1-\hat{G}_{ei}\right)+\gamma\left(p^2+1\right)\left(\alpha+\beta \right) \right]}{2\gamma+\alpha+\beta},
\end{equation}
if $\hat G$ is sufficiently large and negative. Waves at these combinations of {\bf K} and $\Omega$ dominate gamma activity.

\subsection{Transfer Function Due to Resonances}
\label{subsec:transfer_fn}
The correlation analysis of \cite{robinson_visual_2007} approximated the transfer function using only the lowest reciprocal lattice vector $\mathbf{K}$. We generalize that result to include higher order lattice vectors ${\bf K}_j$ that describe finer spatial structure of the OP map, and denote the corresponding frequencies as $\Omega_j$. Then, the transfer function is
\begin{equation}\label{eq_T_ms_k_w_2}
T_{en}(\mathbf{k},\omega)\approx \sum_{\mathbf{K}_{j},\Omega_{j}}\frac{T_{0}(\mathbf{K}_{j},\Omega_{j})}{(\mathbf{k}-\mathbf{K}_{j})^2r_{em}^2+q^2r_{em}^2}\,.
\end{equation}
\begin{equation}\label{eq_T_0}
T_0(\mathbf{k},\omega) = \frac{J_{en}\hat{J}_{em}c_{\mathbf{K}}}
{(1-J_{ei})^2},
\end{equation}
\begin{equation}\label{eq_q}
q^2r_{em}^2=(1-i\omega/\gamma_{em})^2 + \hat{J}_{em}c_{\mathbf{K}}/(1-J_{ei}),
\end{equation}
where $\hat{J}_{em}$ is defined in Eq.~(\ref{eq_J_ab}).
Spatially Fourier transforming Eq.~\eqref{eq_T_ms_k_w_2} then gives
\begin{equation}\label{T_ms_r_w}
T_{en}(\mathbf{r},\omega)\approx \sum_{\mathbf{K}_{j},\Omega_{j}}\left[(2\pi r_{em}^2)^{-1}e^{i\mathbf{K}_{j}\cdot\mathbf{r}}T_0(\mathbf{K}_{j},\Omega_{j})K_0(q\left|\mathbf{r}\right|)\right] ,
\end{equation}
where $K_0$ is a modified Bessel function of the second kind \citep{NIST_handbook}.

\section{Correlation Functions}
\label{sec:correlations_theory}
This section summarizes the use of transfer functions to derive the two-point correlation function between the cortical firing rates measured at two different locations, when the cortex is stimulated at two locations, generalizing the analysis of \cite{robinson_visual_2007} and improving its notation.

\begin{figure}[h!]
\centering
    \includegraphics[width=0.55\textwidth]{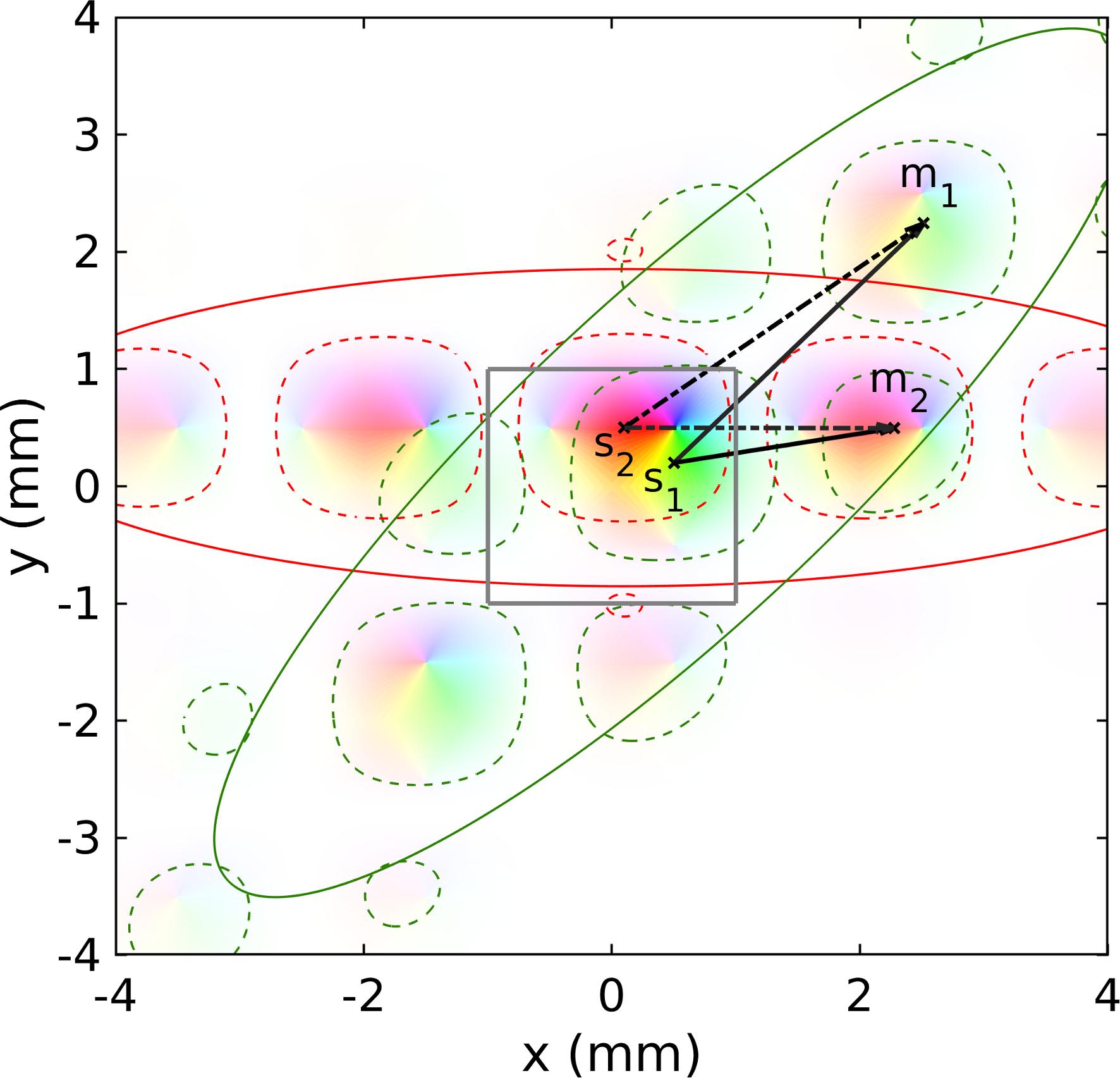}
    \caption{Schematic for deriving the correlation functions, where $\mathbf{s}_{1}$ and $\mathbf{s}_{2}$ denote the stimulus/source points and $\mathbf{m}_{1}$ and $\mathbf{m}_{2}$ are the measurement points. The ellipses in solid green (red) lines indicate the overall shape of the orientation-modulated propagation from $\mathbf{s}_{1}$ ($\mathbf{s}_{2}$), as given by Eq.~\eqref{eq_shape_fn}. The ellipsoids outlined in dotted green (red) lines indicate the patchiness of the propagation along the OP of $\mathbf{s}_{1}$ ($\mathbf{s}_{2}$), with period $k=2\pi/a$.
The solid and dash-dotted arrows denote propagation from $\mathbf{s}_{1}$ and $\mathbf{s}_{2}$, respectively, to $\mathbf{m}_{1}$ and $\mathbf{m}_{2}$.}
\label{fig:schematic_for_CF}
\end{figure}

If the visual cortex receives two uncorrelated and spatially localized inputs at locations $\mathbf{s}_{1}$ and $\mathbf{s}_{2}$. Further, cortical activity is measured at $\mathbf{m}_{1}$ and $\mathbf{m}_{2}$. Figure \ref{fig:schematic_for_CF} shows a schematic of typical spatial locations and OPs involved in deriving the correlation function. The two ellipses in solid green and red, centered at $\mathbf{s}_{1}$ and $\mathbf{s}_{2}$ represent anisotropic propagators $G(\mathbf{r-r'})$ for OPs $\phi(\mathbf{s}_{1}) = 45^{\circ}$ and $\phi(\mathbf{s}_{2}) =  0^{\circ}$. The arrows indicate propagation of neural activity from sources $\mathbf{s}_{j}$ to measurement points $\mathbf{m}_{l}$.

We first derive equations for the neural activities at $\mathbf{m}_{1}$ and $\mathbf{m}_{2}$ due to inputs at $\mathbf{s}_{1}$ and $\mathbf{s}_{2}$. The activity $\Phi$ at $\mathbf{m}_{l}$ can be written as
\begin{equation}
\Phi(\mathbf{m}_{l}, t)=\sum_{j=1,2}\int d^{2}\mathbf{s}_{j} \int d t_{j} T_{en} (\mathbf{m}_{l}, \mathbf{s}_{j}, t-t_{j})\Xi(\mathbf{s}_{j}, t_j),
\end{equation}
where $T_{en}(\mathbf{m}_{l}, \mathbf{s}_{j}, t-t_{j})$ is the transfer function that relates the activities at $\mathbf{m}_{l}$ and time $t$ to the stimulus $\Xi$ at $\mathbf{s}_{j}$ and time $t_j$.

\citet{robinson_visual_2007} approximated a spatially localized input $\Xi(\mathbf{s}_{j},\omega)$ as
\begin{equation}\label{eq_Xi_s1_spatial}
\Xi(\mathbf{s}_{j},\omega)=A_{j}(\omega)\updelta(\mathbf{r}-\mathbf{s}_{j})e^{i\psi(\mathbf{r}, \omega)},
\end{equation}
whence
\begin{equation}\label{eq_Xi_s1_fourier}
\Xi(\mathbf{k}, \omega)=A_{j}(\omega)e^{-i\mathbf{k}\cdot\mathbf{s}_{j}}e^{i\psi(\mathbf{s}_{j}, \omega)},
\end{equation}
where the real quantities $A_{j}(\omega)$ and $\psi(\mathbf{s}_{j},t_{j})$ are the amplitude and the phase of the input at $\mathbf{s}_{j}$. We then find
\begin{equation}\label{eq_phi_1}
\Phi(\mathbf{k}, \omega)= T_{en} (\mathbf{k}, \omega)\sum_{j=1,2} A_{j}(\omega)e^{-i\mathbf{k}\cdot\mathbf{s}_{j}}e^{i\psi(\mathbf{s}_{j}, \omega)}.
\end{equation}

The two-point correlation function between $\mathbf{m}_{1}$ and $\mathbf{m}_{2}$ is \citep{robinson_visual_2007}
\begin{equation}
C(\mathbf{m}_{1}, \mathbf{m}_2, \tau)= \left \langle  \Phi_{1}(\mathbf{m}_{1},t'+\tau)\Phi_{2}(\mathbf{m}_{2}, t')\right \rangle\,,
\end{equation}
where $\tau=t-t'$, and the angle brackets refer to the averages over $t'$ and over the phase of the inputs. A  Fourier transform and integration over $t'$ achieves the averaging \cite{robinson_visual_2007} to yield
\begin{eqnarray}\label{eq_corr_1}
C(\mathbf{m}_{1}, \mathbf{m}_{2}, \tau)&=&\int d t' \int \frac{d \omega }{2\pi}\int\frac{d \omega'}{2\pi}\int \frac{d^{2} \mathbf{k}}{(2\pi)^{2}}\int \frac{d^{2} \mathbf{k'}}{(2\pi)^{2}}\nonumber\\
\quad &\times& e^{-i\omega(t'+ \tau)+i\omega't'+i\mathbf{k}\cdot \mathbf{m}_{1}-i\mathbf{k'}\cdot \mathbf{m}_{2}} \left\langle  \Phi_{1}(\mathbf{k},\omega)\Phi_{2}^*(\mathbf{k'}, \omega')\right \rangle,\\
&=&\int \frac{d \omega }{2\pi}\int \frac{d^{2} \mathbf{k}}{(2\pi)^{2}}\int \frac{d^{2} \mathbf{k'}}{(2\pi)^{2}}\nonumber\\
      &\times& e^{-i\omega \tau +i\mathbf{k}\cdot \mathbf{m}_{1}-i\mathbf{k'}\cdot \mathbf{m}_{2}}\left\langle  \Phi_{1}(\mathbf{k},\omega)\Phi_{2}^*(\mathbf{k'}, \omega)\right \rangle.
\end{eqnarray}
Substituting Eq.~\eqref{eq_phi_1} into Eq.~\eqref{eq_corr_1}, and taking the inverse Fourier transform then gives
\begin{eqnarray}
\label{eq_corr_m1_m2_3}
C(\mathbf{m}_{1}, \mathbf{m}_{2}, \tau)&=& \Bigg \langle \int \frac{d \omega }{2\pi} e^{-i\omega \tau} \nonumber\\
 &\times& \left [ T_{en} (\mathbf{m}_{1}-\mathbf{s}_{1}, \omega)A_{1}(\omega)e^{i\psi(\mathbf{s}_{1}, \omega)} + T_{en} (\mathbf{m}_{1}-\mathbf{s}_{2}, \omega)A_{2}(\omega)e^{i\psi(\mathbf{s}_{2}, \omega)}\right ]\nonumber\\
      &\times& \left [T_{en}^{*} (\mathbf{m}_{2}-\mathbf{s}_{1}, \omega)A_{1}(\omega)e^{-i\psi(\mathbf{s}_{1}, \omega)} + T_{en} ^{*}(\mathbf{m}_{2}-\mathbf{s}_{2}, \omega)A_{2}(\omega)e^{-i\psi(\mathbf{s}_{2}, \omega)} \right ] \Bigg \rangle,
\end{eqnarray}
where the angle brackets now denote the average over the phases at $\mathbf{s}_{1}$ and $\mathbf{s}_{2}$.
If the phases of the inputs are random and uncorrelated,
\begin{equation}
\left \langle e^{i\psi(\mathbf{s}_{1}, \omega)}e^{i\psi(\mathbf{s}_{2}, \omega)}\right \rangle = \updelta^2(\mathbf{s}_{1}-\mathbf{s}_{2}),
\end{equation}
so the cross terms between $\mathbf{s}_{1}$ and $\mathbf{s}_{2}$ in Eq.~\eqref{eq_corr_m1_m2_3} are zero and
\begin{eqnarray}\label{eq_corr_m1_m2_final}
C(\mathbf{m}_{1}, \mathbf{m}_{2}, \tau)&=& \int \frac{d \omega }{2\pi} e^{-i\omega \tau} \nonumber\\
 &\times& \left\{ T_{en} (\mathbf{m}_{1}-\mathbf{s}_{1}, \omega) T_{en}^{*} (\mathbf{m}_{2}-\mathbf{s}_{1}, \omega)|A_{1}(\omega)|^2\right. \nonumber\\
 &+& \left. T_{en} (\mathbf{m}_{1}-\mathbf{s}_{2}, \omega) T_{en}^{*} (\mathbf{m}_{2}-\mathbf{s}_{2}, \omega)|A_{2}(\omega)|^2 \right \},
\end{eqnarray}
which is the sum of the correlations due to the two stimuli taken separately.

Finally, substituting Eq.~(\ref{T_ms_r_w}) into Eq.~(\ref{eq_corr_m1_m2_final}), assuming inputs at different ${\bf K}_j$ are uncorrelated, and letting $\left|A_{1}(\omega)\right|=\left|A_{2}(\omega)\right|=1$ for simplicity, gives
\begin{eqnarray}\label{eq_corr_m1_m2_numerical}
C(\mathbf{m}_{1}, \mathbf{m}_{2}, \tau)&=& (2\pi r_{em}^2)^{-1}\int \frac{d \omega }{2\pi} e^{-i\omega \tau} \nonumber\\
 &\times& \sum_{\mathbf{K}_{j},\Omega_{j}}\left \{ \left[e^{i\mathbf{K}_{j}\cdot(\mathbf{m}_1-\mathbf{s}_{1})}T_0(\mathbf{K}_{j},\Omega_{j})K_0(q\left |\mathbf{m}_{1}-\mathbf{s}_{1}\right |)\right] \right.\nonumber\\
 &\times& \left .\left[e^{i\mathbf{K}_{j}\cdot(\mathbf{m}_2-\mathbf{s}_{1})}T_0(\mathbf{K}_{j},\Omega_{j})K_0(q\left |\mathbf{m}_{2}-\mathbf{s}_{1}\right |)\right]^{*} \right \}\nonumber\\
 &+&\left \{\left[e^{i\mathbf{K}_{j}\cdot(\mathbf{m}_{1}-\mathbf{s}_{2})}T_0(\mathbf{K}_{j},\Omega_{j})K_0(q\left |\mathbf{m}_{1}-\mathbf{s}_{2}\right |)\right] \right.\nonumber\\
 &\times& \left .\left[e^{i\mathbf{K}_{j}\cdot(\mathbf{m}_2-\mathbf{s}_{2})}T_0(\mathbf{K}_{j},\Omega_{j})K_0(q\left |\mathbf{m}_{2}-\mathbf{s}_{2}\right |)\right]^{*}\right \}  \,.
\end{eqnarray}
Some general aspects of Eq.~\eqref{eq_corr_m1_m2_numerical} are that the correlations fall off on a characteristic spatial scale of $({\rm Re}q)^{-1}$ because $K_0(z)\sim \exp(-z)$ at large $z$ in the right half plane. For the same reason, there is an oscillation with spatial frequency of ${\rm Im}q$, while resonances in $T_0$ select dominant temporal frequencies in the correlations.

\section{Patchy Propagation}
\label{sec:propagation}
\cite{robinson_visual_2007}showed that the gamma response can be approximated as a sum of resonant responses at various ${\bf K}_j$. He further analyzed a spatially 1D system by approximating the contributions of these poles as Gaussians in $k-\omega$ space. This yielded patchy propagation with a Gaussian envelope as a function of distance, which explained a number of gamma correlation properties.

Here we generalize the analysis of \cite{robinson_visual_2007} to the spatially 2D cortex and to allow for the spatial anisotropy of the envelope of patchy connections, which extend furthest along a direction corresponding to the orientation of the source OP. We quantify the patchy propagation via the coefficients $c_{\bf K}$ in Eq.~\eqref{eq_general_prop}. \cite{robinson_visual_2007} previously approximated the spatial propagator in 1D as a Gaussian function. The propagation was assumed to be isotropic with its patchiness described as $\cos(\mathbf{K} x)$, from which is formed by a pair of complex conjugated coefficients $c_{\bf +K}$ and $c_{\bf -K}$, $\bf K$ is the lowest reciprocal lattice vector. However, in 2D, patches of neurons with similar feature preference are preferentially connected, with connections \citep{bressloff_functional_2003,Gilbert_patchy_connectn,lund_anatomical_2003,muir_embedding_2011}, concentrated toward an axis corresponding to their OP angle \citep{bosking_orientation_1997,Malach_horizontal_connect, Sincich_oriented_connct}. To model this overall modulation of the anisotropic propagation, we approximate the spatial propagator at each point and Fourier transform it to obtain a set of coefficients $c_{\mathbf{K}_{j}}$, where $\mathbf{K}_{j}$ corresponding to the reciprocal lattice vectors. These coefficients $c_{\mathbf{K}_{j}}$ are used to calculate the transfer function $T_{en}$ described by Eqs~\eqref{eq_T_0} and \eqref{T_ms_r_w}.

\begin{figure}[h]
\begin{multicols}{2}
\centering
\includegraphics[width=0.35\textwidth]{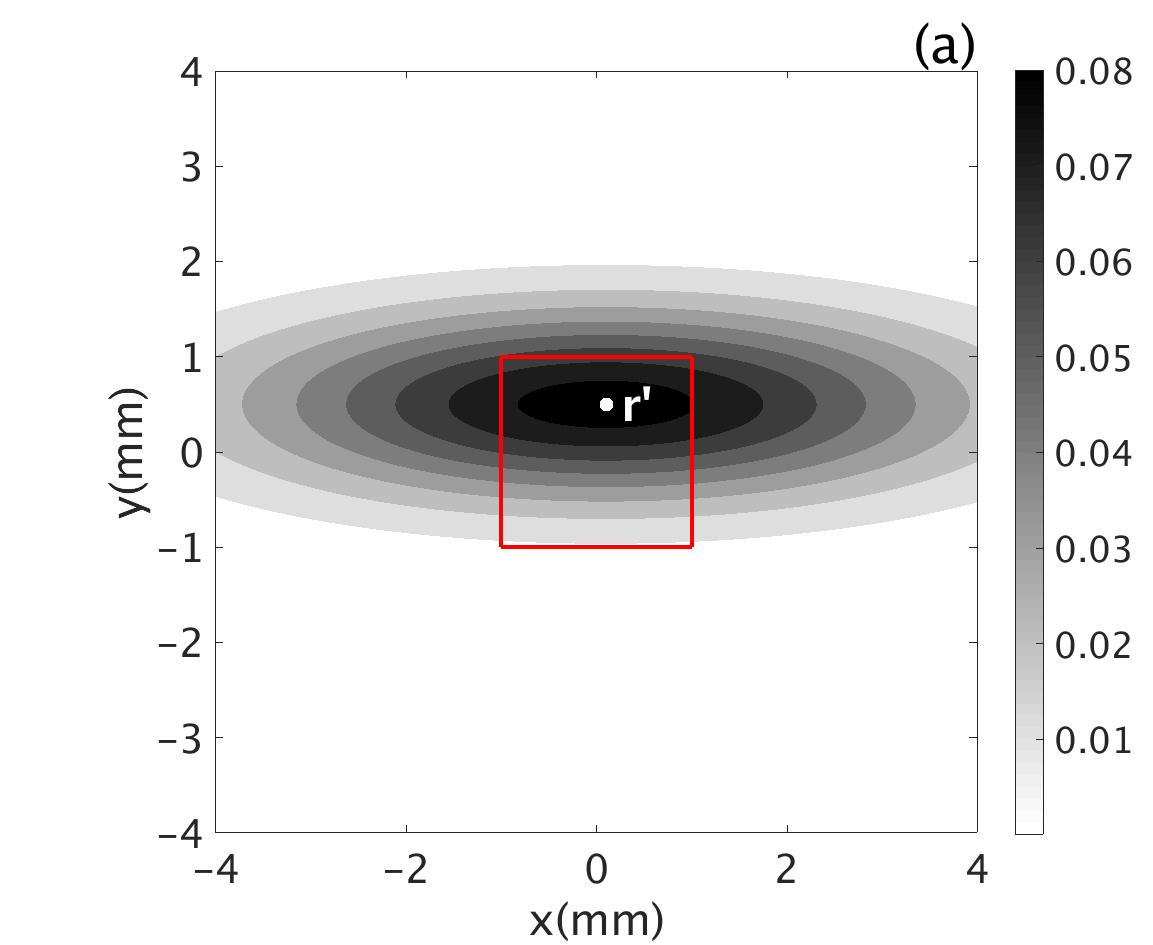}\\
\includegraphics[width=0.35\textwidth]{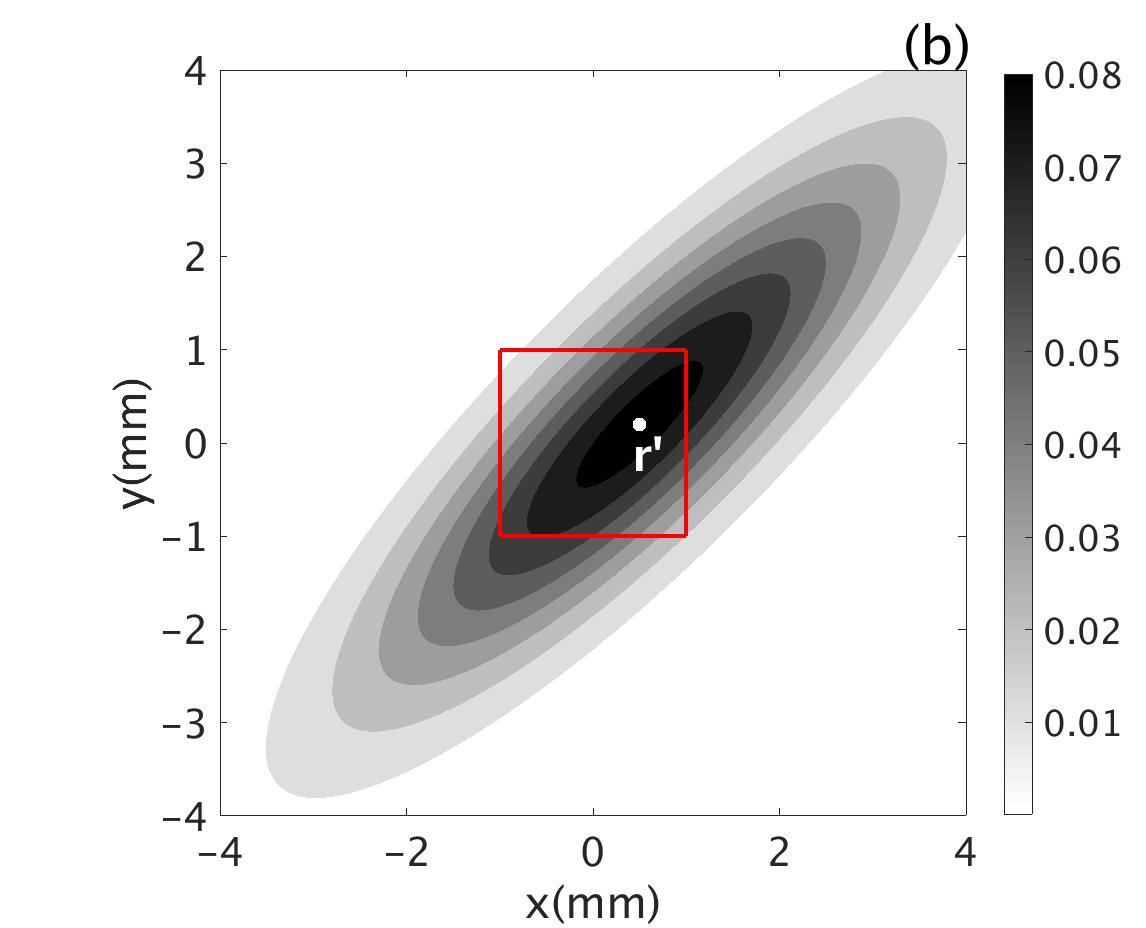}
\end{multicols}
\caption{Plots of Eq.~\eqref{eq:shape_fn_strength} with the central unit cell outlined in red; the color bar shows values of $G(\mathbf{r}-\mathbf{r'})$. (a) OP $=0^{\circ}$. (b) OP $=45^{\circ}$.}
\label{fig:elliptic_gaussian}
\end{figure}

A reasonable approximation to the envelope of the patchy connections that emerge from a particular point ${\bf r}'$ is an elliptic Gaussian whose long axis is oriented at the local OP $\phi$ at $\mathbf{r'}$. If $\mathbf{r'}=(x',y')$ and $\mathbf{r}=(x,y)$, we have
\begin{equation}
G(\mathbf{r}-\mathbf{r'})= \frac{1}{2\pi \sigma_{x} \sigma_{y}}\exp \left [-\frac{1}{2}\left(\frac{x_{g}^{2}}{\sigma_{x}^{2}}+\frac{y_{g}^{2}}{\sigma_{y}^{2}}\right)\right ],
\label{eq:shape_fn_strength}
\end{equation}
where
\begin{equation}
x_{g}=(x-x')\cos\left[\phi(x',y') \right]+(y-y')\sin\left[\phi(x',y') \right],
\end{equation}
\begin{equation}
y_{g}=-(x-x')\sin\left[\phi(x',y') \right]+(y-y')\cos\left[\phi(x',y') \right].
\end{equation}
where $\sigma_{x} =$ \SI{2.6}{mm} and $\sigma_{y}=$ \SI{0.7}{mm} are the spatial ranges along the preferred $x_{g}$ and orthogonal $y_{g}$ directions, with values chosen to match the experimental findings in tree shrew by \cite{bosking_orientation_1997}.
Figures~\ref{fig:elliptic_gaussian}(a) and (b) show contour plots of $G(\mathbf{r}-\mathbf{r'})$ for OPs of 0$^{\circ}$ and 45$^{\circ}$, respectively and source points $\mathbf{r'}$ within a central unit cell [see Fig.~\ref{fig:op_map}(c)].

\begin{figure}[h]
\begin{multicols}{2}
\centering
\includegraphics[width=0.35\textwidth]{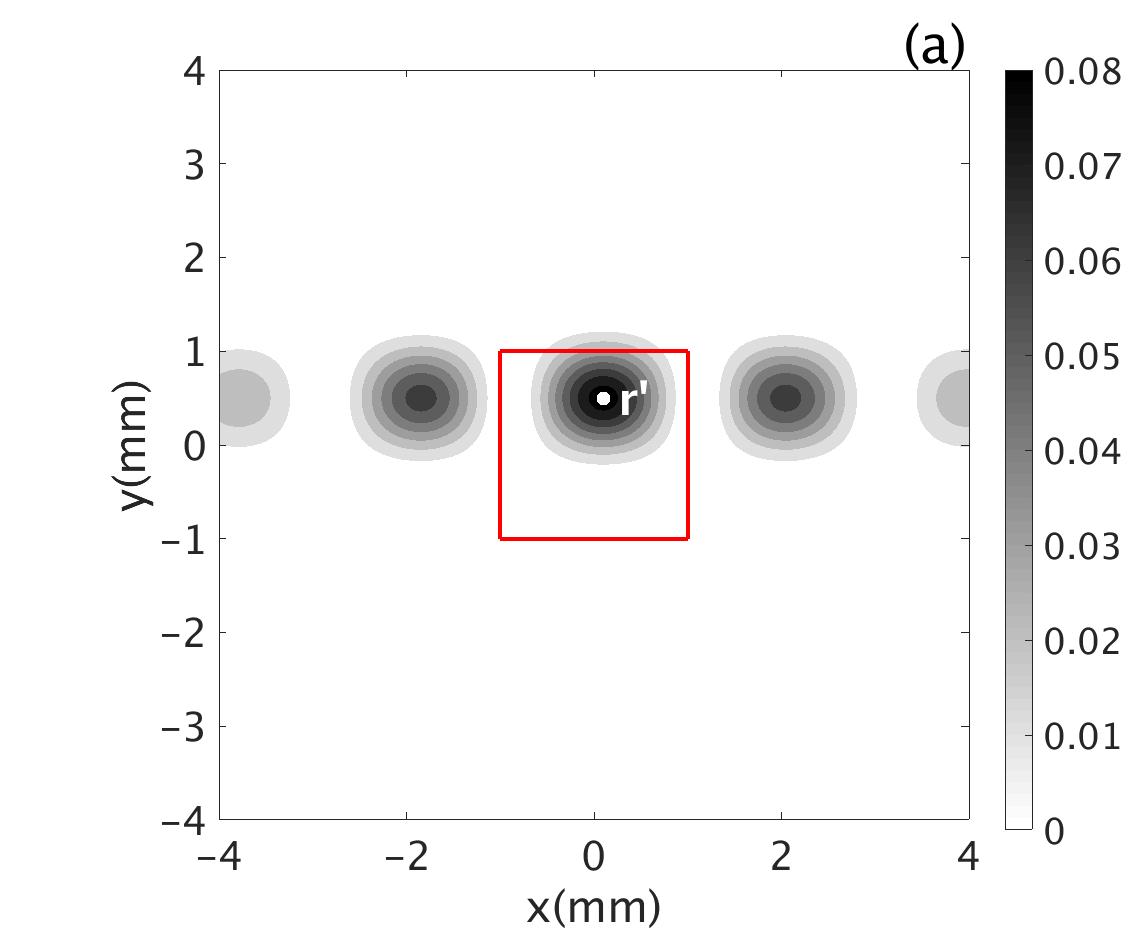}\\
\includegraphics[width=0.35\textwidth]{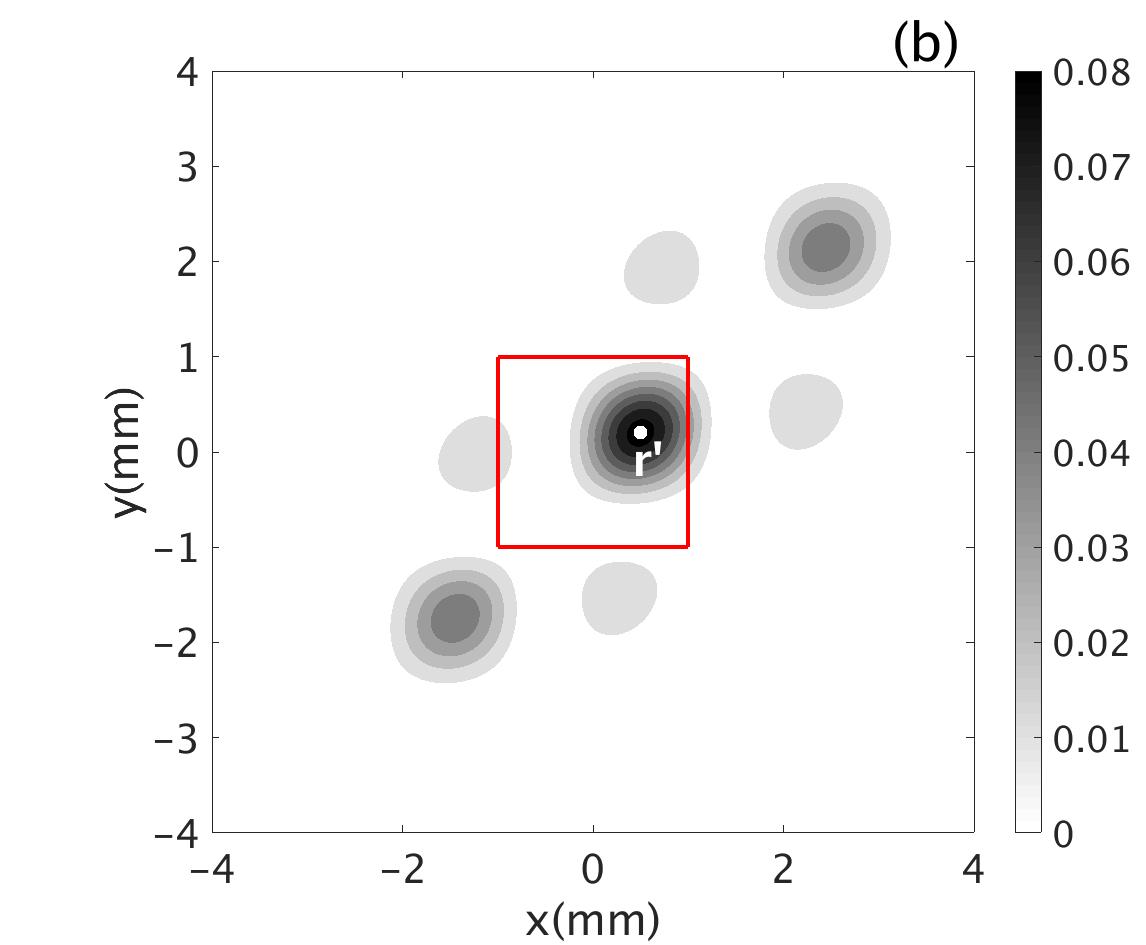}
\end{multicols}
\caption{Patchy propagator $G(\mathbf{r},\mathbf{r'})$ in Eq.~\eqref{eq_shape_fn} with the central unit cell outlined in red containing the source point $\mathbf{r}'$. The color bar shows the values of $G(\mathbf{r},\mathbf{r'})$. (a) $\phi(\mathbf{r'})=0^{\circ}$. (b) $\phi(\mathbf{r'})=45^{\circ}$. }
\label{fig:elliptic_gaussian_on_cos}
\end{figure}

Patchy propagation is modulated with spatial period $k=2\pi/a$ parallel and orthogonal to OD columns, where $a\approx$ 2 mm is the width of the unit cell. To incorporate this modulation, we multiply the oriented elliptic Gaussian function by a product of cosine functions that reflect this periodicity. This gives an  approximate propagator profile of the form 
\begin{eqnarray}\label{eq_shape_fn}
G(\mathbf{r}-\mathbf{r'})&= &\frac{1}{2\pi \sigma_{x} \sigma_{y}}\exp \left [-\frac{1}{2}\left(\frac{x_{g}^{2}}{\sigma_{x}^{2}}+\frac{y_{g}^{2}}{\sigma_{y}^{2}}\right)\right ]\nonumber\\
&\times& \left\lbrace\cos[k_{x}(x-x')]+1\right\rbrace \left\lbrace\cos[k_{y}(y-y')]+1\right\rbrace,
\end{eqnarray}
where $k_{x}=k_{y}=2\pi/a$. We use this functional form to generalize the 1D cosine-modulated Gaussian form of \cite{robinson_visual_2007} to represent the propagator of a given resonance in the 2D anisotropic case. Figures~\ref{fig:elliptic_gaussian_on_cos} (a) and (b) show the resulting propagators for $\phi(\mathbf{r'})=0^{\circ}, 45^{\circ}$, with $\sigma_{x}=$ \SI{2.6}{mm} and $\sigma_{y}=$ \SI{0.7}{mm}. For both cases, when $\mathbf{r}-\mathbf{r'} <$ \SI{0.5}{mm} the underlying neurons respond to the stimulus, regardless of OP.

\begin{figure}[h]
\begin{multicols}{2}
\centering
\includegraphics[width=0.35\textwidth]{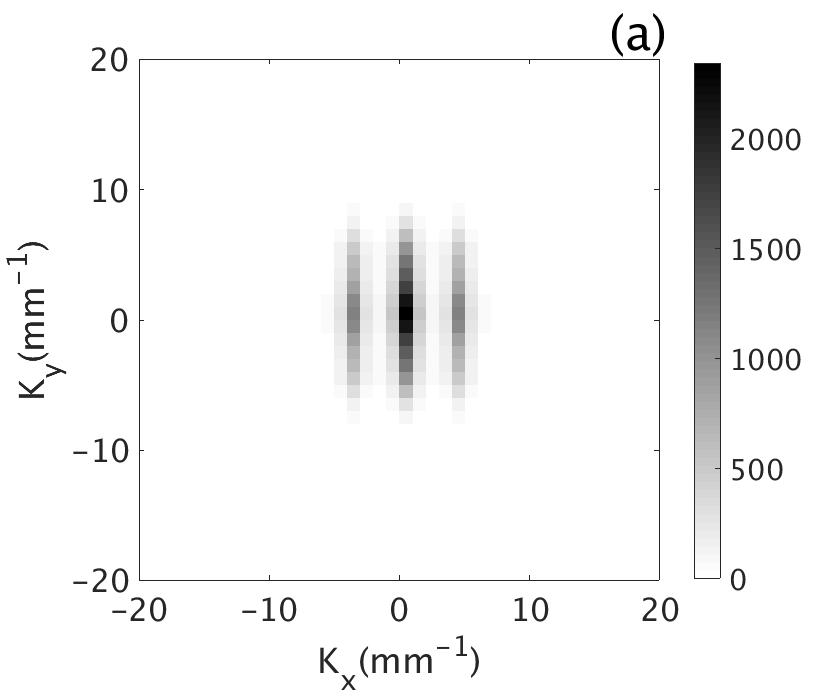}\\
\includegraphics[width=0.35\textwidth]{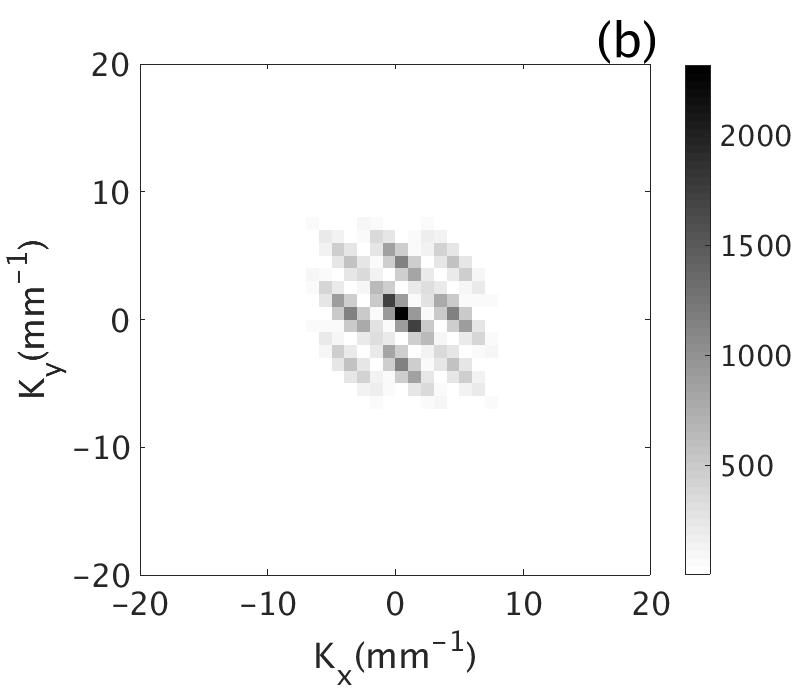}
\end{multicols}
\caption{Fourier coefficients of $G(\mathbf{r},\mathbf{r'})$ in Eq.~\eqref{eq_shape_fn}. Each pixel-like square represents one $c_{\mathbf{K}_{j}}$. The color bar shows the magnitudes of the coefficients. (a) $\phi(\mathbf{r'})=0^{\circ}$. (b) $\phi(\mathbf{r'})=45^{\circ}$. }
\label{fig:FT_gaussian_on_cos}
\end{figure}

After performing a 2D Fourier transform on the propagators shown in Figure~\ref{fig:elliptic_gaussian_on_cos}, the coefficients $c_{\mathbf{K}_{j}}$ are illustrated in Figure \ref{fig:FT_gaussian_on_cos} .  Both two sets of coefficients do not have high frequency components. In later section, we choose a fraction of $c_{\mathbf{K}_{j}}$, which preserves the basic spatial propagation structure, for evaluating the transfer function.

\section{Spatiotemporal properties of the correlation function}\label{sec:temporal_corr_properties}
Here, we first explore the temporal properties of the correlation function in Eq.~\eqref{eq_corr_m1_m2_numerical}. Then we explore its spatial properties with a single input. Lastly, we examine the spatial correlation in the case of two input sources.

In all the cases described below, the correlation is calculated by numerically evaluating Eq.~\eqref{eq_corr_m1_m2_numerical} and locating $\mathbf{m}_{1}$, $\mathbf{m}_{2}$, $\mathbf{s}_{1}$, and $\mathbf{s}_{2}$ under different conditions. These conditions include using different optimal OPs for the measurement points and source points, and varying the distances between the measurement points. The results are presented in Fig.~\ref{fig:mulit_corr_fn}. All correlations are normalized such that $C(\mathbf{m}_{1}, \mathbf{m}_{2}, \tau) = 1$ when $\mathbf{s}_{1} = \mathbf{s}_{2}$, and $\mathbf{m}_{1} = \mathbf{m}_{2}$ are placed very close to the sources. Table \ref{brain_para} summarizes the parameters we use for the calculations.
\begin{table}[h!]
\centering
\caption{Nominal EMIRS model parameters.}
\label{brain_para}
\begin{tabular}{llll}
\hline\hline
Synaptodendritic rates                     & $\alpha_{em}$, $\alpha_{es}$, $\alpha_{ei}$ & 80    & s$^{-1}$                     \\
                                              & $\beta_{em}$, $\beta_{es}$, $\beta_{ei}$    & $800$   & s$^{-1}$                     \\
Projection Range                              & $r_{em}$                                   & $2$     & mm                                       \\
                                              & $r_{ei}$                                   & $0.2$   & mm                                       \\
                                              & $r_{es}$                                   & $0.3$   & mm                                       \\
Damping rates                                 & $\gamma_{em}$                              & $500$   & s$^{-1}$                     \\
                                              & $\gamma_{ei}$                              & $1500$  & s$^{-1}$                     \\
Gains                                         & $G_{es}$                                  & $1.7$   &                                          \\
                                              & $G_{em}$                                  & $6.9$  &                                          \\
                                              & $G_{ei}$                                 & $-15.0$ &                                          \\
$dQ_e/dV_e$ & $\rho_{e}$                                 & $4200$  & ${\rm V}^{-1}$s$^{-1}$\\
\hline
\end{tabular}
\end{table}

\subsection{Temporal Correlation Properties}
\label{subsec:temporal_corr}
\begin{figure}[ht!]
\begin{multicols}{2}%
\centering
\includegraphics[width=.39\textwidth]{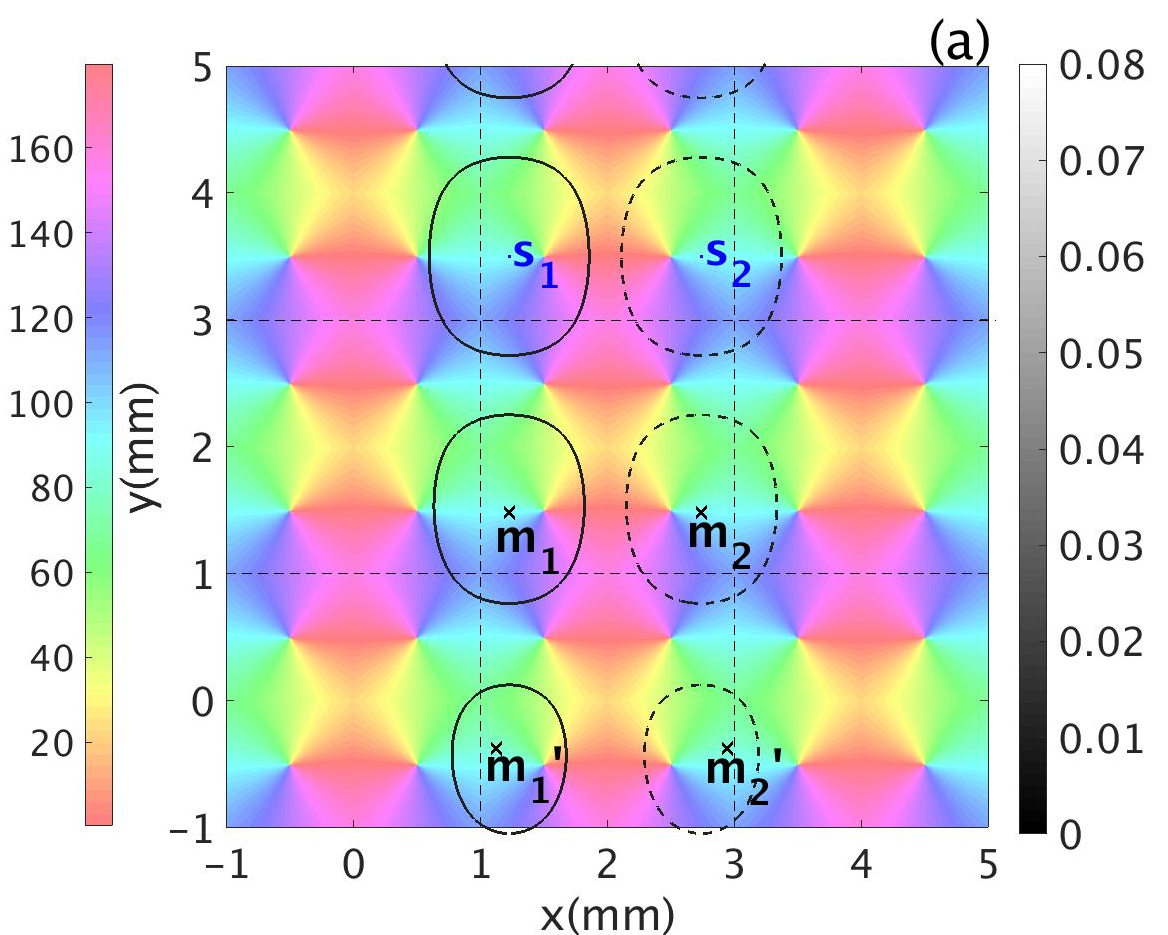}
\includegraphics[width=.39\textwidth]{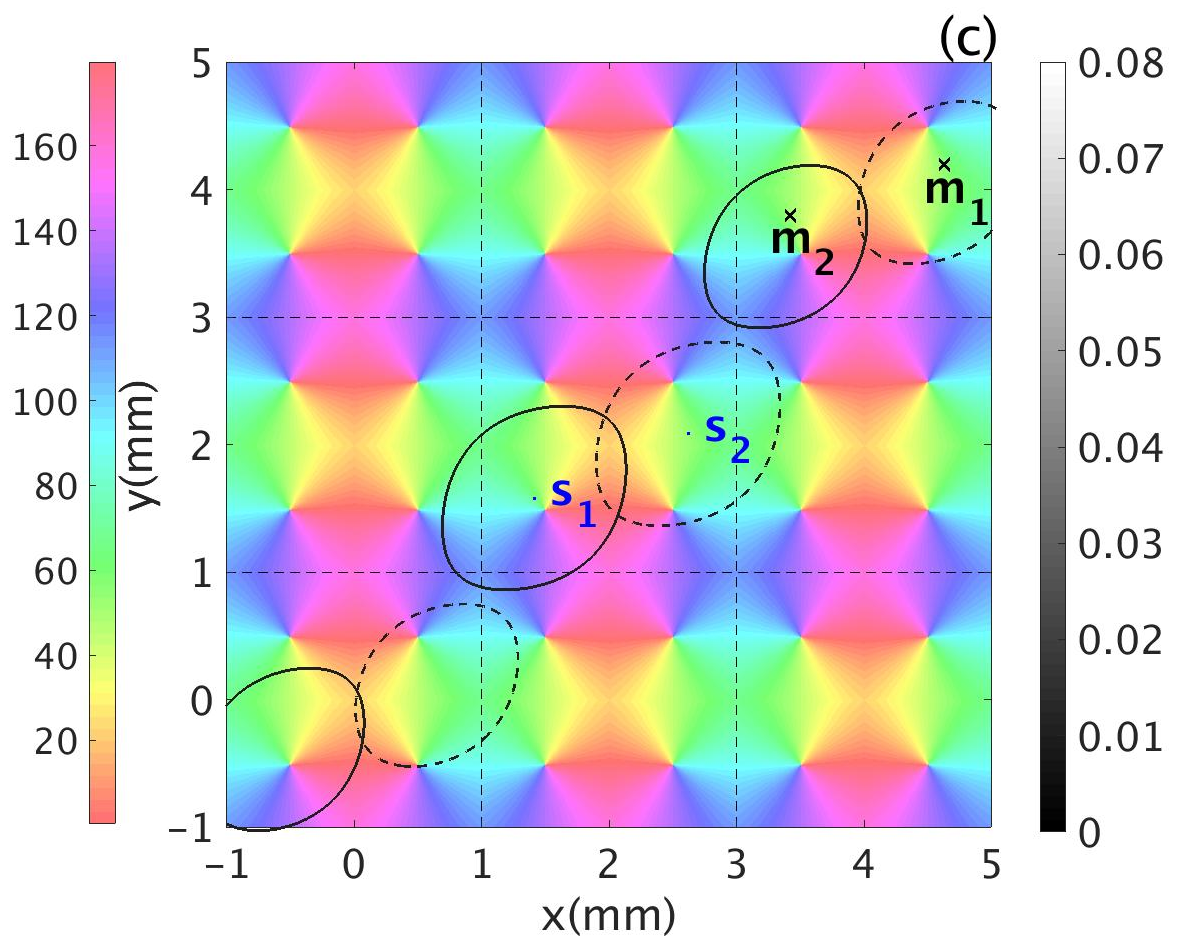}
\includegraphics[width=.39\textwidth]{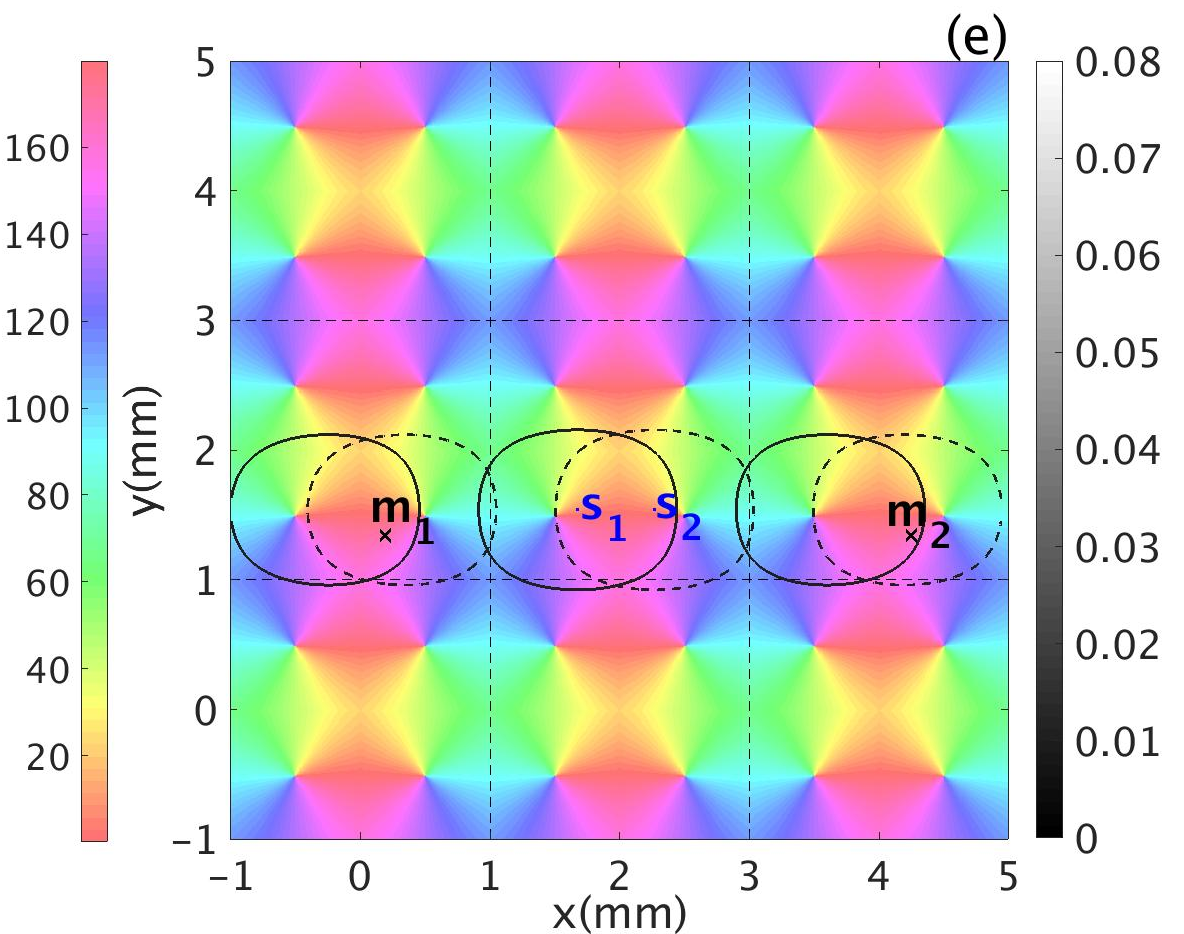}
\includegraphics[width=.39\textwidth]{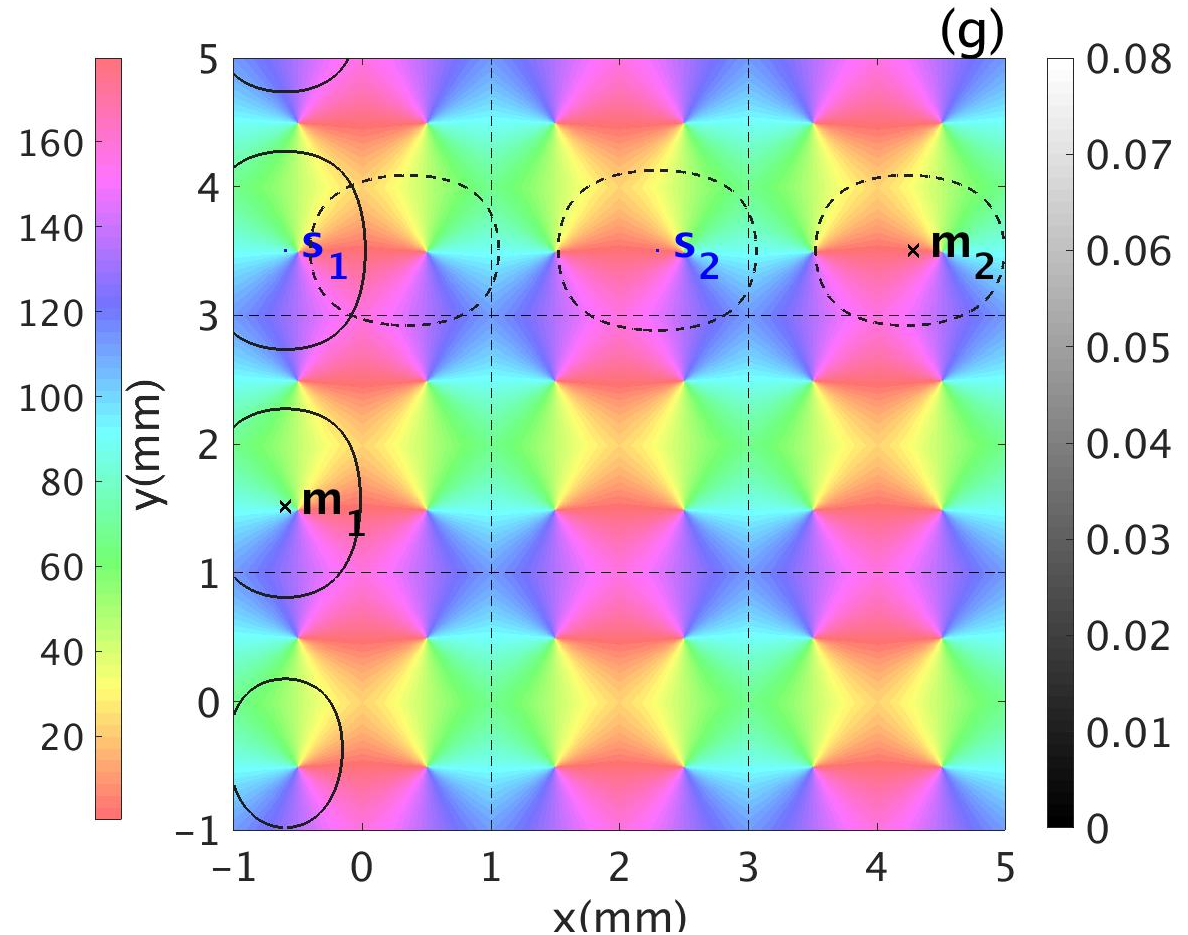}
\includegraphics[width=.32\textwidth]{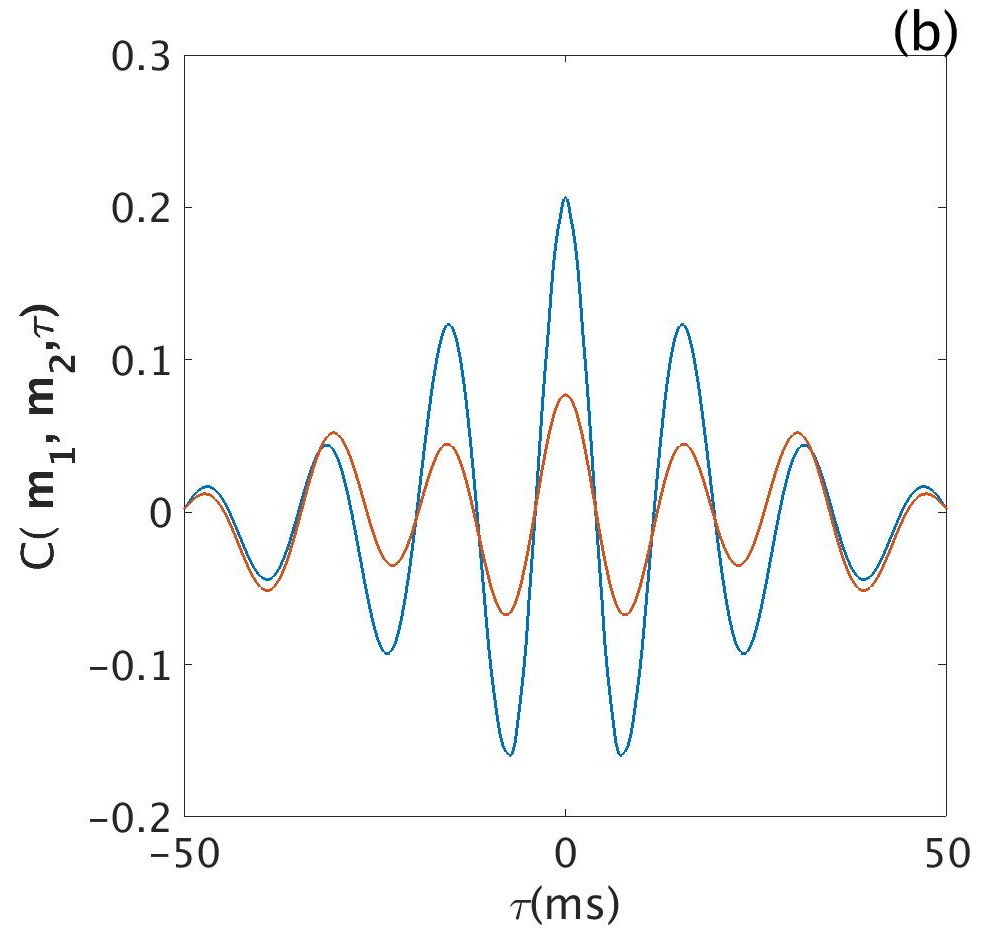}
\includegraphics[width=.32\textwidth]{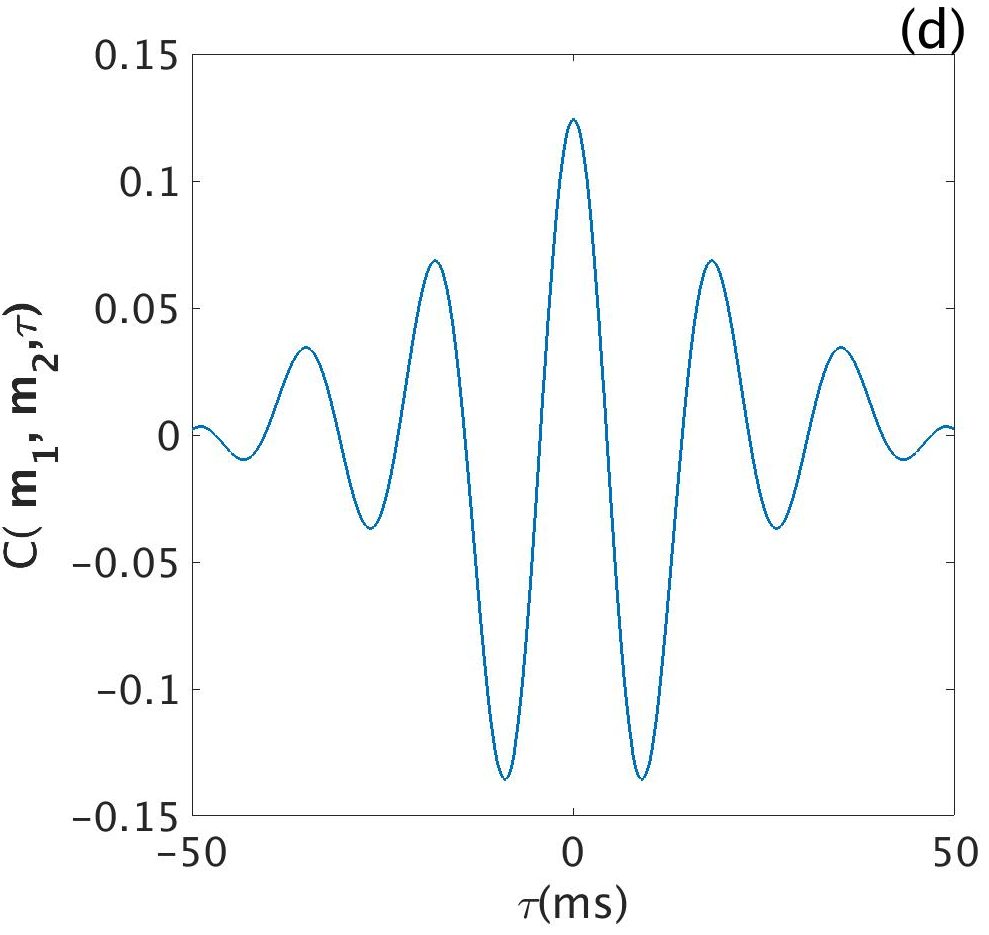}
\includegraphics[width=.32\textwidth]{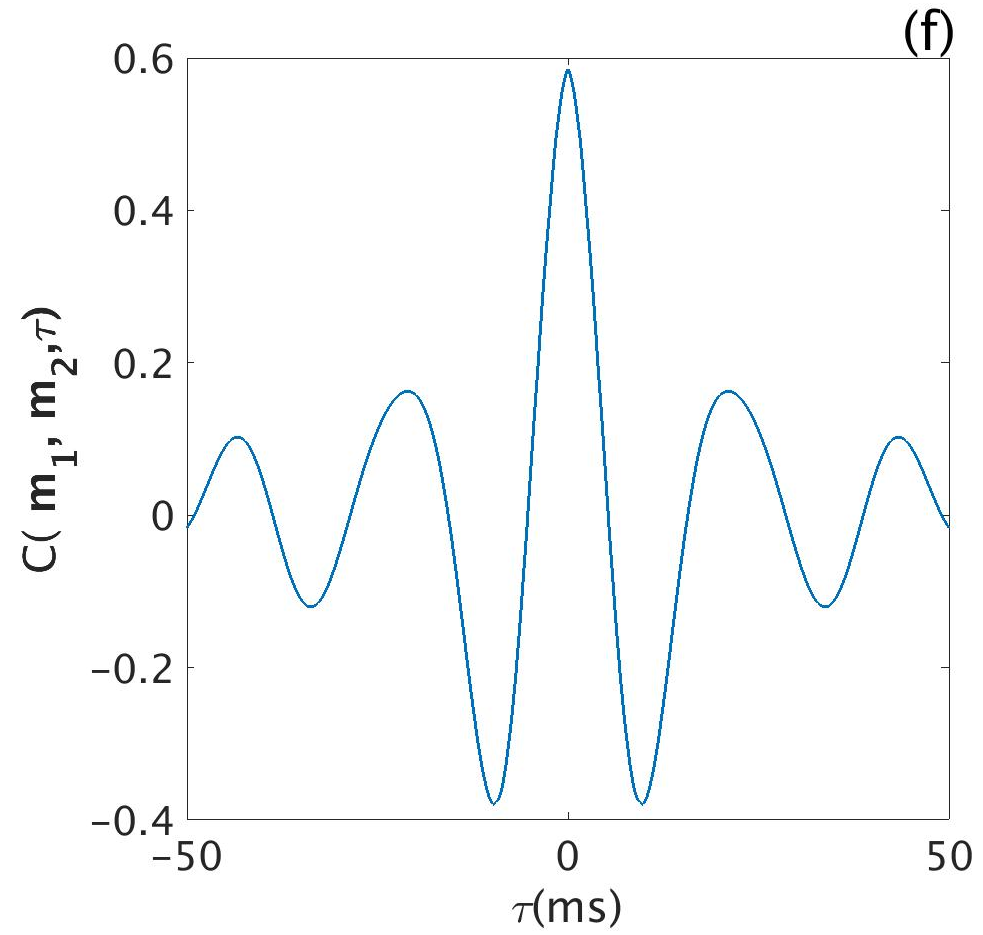}
\includegraphics[width=.32\textwidth]{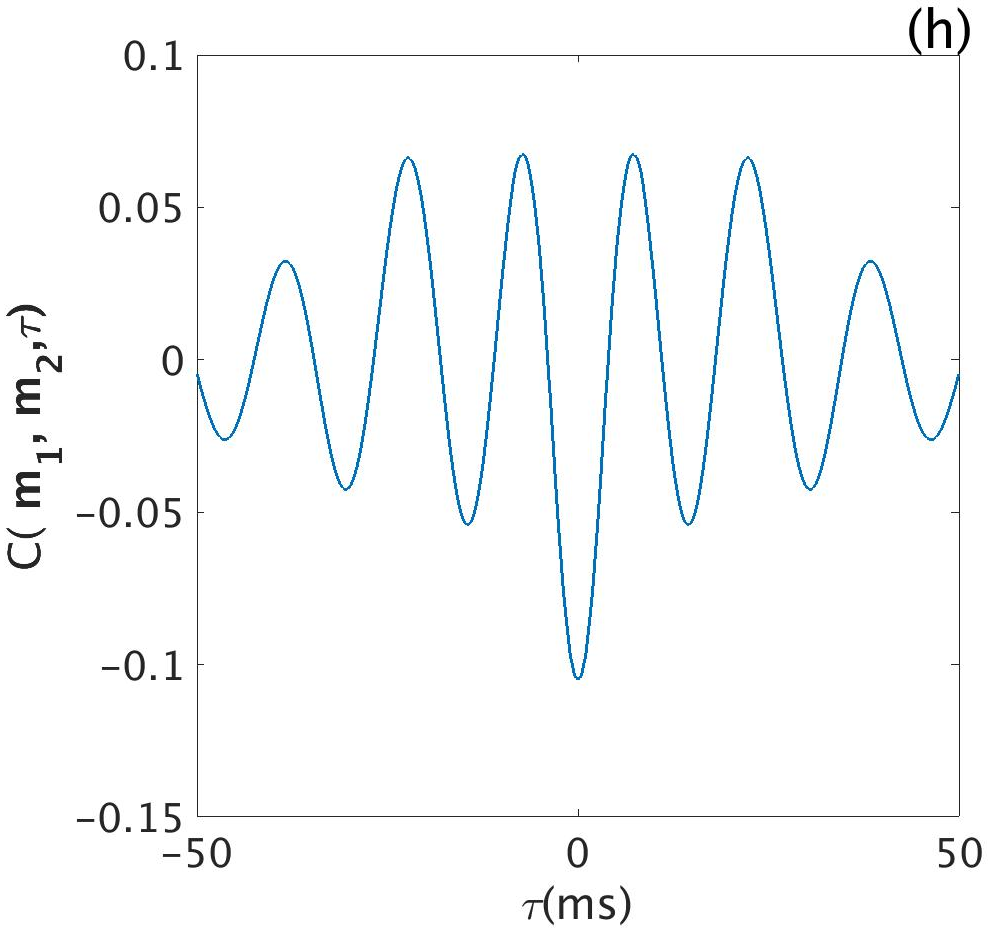}
\end{multicols}
\caption{Temporal correlations. (a) Locations of measurement points $\mathbf{m}_{1}$ and $\mathbf{m}_2$, and source points $\mathbf{s}_{1}$, and $\mathbf{s}_{2}$ within 9 unit cells in V1. All points have OP of 90$^\circ$. The color bar shows the OP and the gray contours show the strength of propagators given by Eq.~\eqref{eq_shape_fn} with solid and dashed curves for propagation from ${\bf s}_1$ and ${\bf s}_2$, respectively. (b) Temporal correlations. Blue curve shows $C(\mathbf{m}_{1}, \mathbf{m}_{2}, \tau)$, while the orange curve shows  $C(\mathbf{m}_{1}',\mathbf{m}_{2}',\tau)$. (c) As for (a) but with all points have OP of 45$^\circ$. (d) Temporal correlation for (c). (e) As for (a) but with all points have OP of 0$^\circ$. (f) Temporal correlation for (e). (g) As for (a) but with OP of the sources are orthogonal. Furthermore, OP at $\mathbf{s}_{1}$ is optimal for $\mathbf{m}_{1}$ whereas OP at $\mathbf{s}_{2}$ is optimal to $\mathbf{m}_2$. (h) Temporal correlation for (g).}
\label{fig:mulit_corr_fn}
\end{figure}

In Fig.~\ref{fig:mulit_corr_fn}(a), we illustrate the temporal correlations evoked by binocular stimulation when $\mathbf{s}_{1}$ and $\mathbf{s}_{2}$ have the same OP $\phi \mathbf{(s)} = 90^\circ$ and $\mathbf{s}_{1}$ and $\mathbf{s}_{2}$ are located in the same unit cell, but in different OD columns. The strength of propagation of neural signals from two sources is indicated by contour lines of Eq.~\eqref{eq_shape_fn}; the propagation is predominantly parallel in this case.

The points $\mathbf{m}_{1}$ and $\mathbf{m}_{2}$ are located in a different unit cell to the source points; are approximately \SI{2}{mm} away from each other; and are located at approximately \SI{2}{mm} from their respective collinear source points, the OPs at $\mathbf{m}_{1}$ and $\mathbf{m}_{2}$ are also $90^\circ$. We have also placed additional measurement points $\mathbf{m}_{1}'$ and $\mathbf{m}_{2}'$, with the same OP as $\mathbf{m}_{1}$ and $\mathbf{m}_{2}$, but are approximately \SI{4}{mm} from the sources.

Figure \ref{fig:mulit_corr_fn}(b) shows the temporal correlation functions $C(\mathbf{m}_{1}, \mathbf{m}_{2}, \tau)$ and $C(\mathbf{m}_{1}', \mathbf{m}_{2}',\tau)$. Both
oscillate at around \SI{64}{Hz}, in the gamma range. Furthermore, each has a peak centered at $\tau=0$, so the neural activities at $\mathbf{m}_{1}$ and $\mathbf{m}_{2}$, $\mathbf{m}_{1}'$ and $\mathbf{m}_{2}'$, are synchronized. The time for their envelopes to decrease to $1/e$ ($\sim 30\%$) of the peak value is $\approx$ \SI{18}{ms}. However, when the measurement points are placed further away from the sources, the correlation at $\tau=0$ becomes weaker, as seen by comparing the two curves.

Figure~\ref{fig:mulit_corr_fn}(c) shows a case for which the OP of all sources and measurement points is equal (at $45^\circ$). Figure~\ref{fig:mulit_corr_fn}(d) shows that the resulting correlation also has a central peak at zero time-lag, oscillates in the gamma band at $\sim$\SI{55}{Hz}, and its envelope decreases by $1/e$ at $\tau\approx$ \SI{21}{ms}.

In order to explore the correlation properties between OD columns, we place all the source points and measurement points co-linearly with OP $=0^\circ$ in Figure \ref{fig:mulit_corr_fn}(e). Synchronized activities at $\mathbf{m}_{1}$ and $\mathbf{m}_{2}$ are shown by the center peak at $\tau=0$ in Fig.~\ref{fig:mulit_corr_fn}(f). This correlation also exhibits gamma band oscillation at $\approx$ \SI{50}{Hz}, and the decrease by $1/e$ from the peak happens at $\approx$ \SI{21}{ms}. One thing worth to be mentioned here is that the correlation strength due to inter-columnar connection shown in Fig.~\ref{fig:mulit_corr_fn}(f), is stronger than the intra-column connection in Fig.~\ref{fig:mulit_corr_fn}(b).

To further investigate the correlation properties, Fig.~\ref{fig:mulit_corr_fn}(g) shows a case in which the two measurement sites have orthogonal OPs, and so does the sources: the OP at $\mathbf{s}_{1}$ and $\mathbf{m}_{1}$ is $90^\circ$, while at $\mathbf{s}_{2}$ and $\mathbf{m}_{2}$ it is $0^\circ$. The distance between the two measurement points is around \SI{5.5}{mm}. In this case, $\mathbf{s}_{1}$ tends to evoke strong response at $\mathbf{m}_{1}$, but not at $\mathbf{m}_{2}$. This introduces an anticorrelation between $\mathbf{m}_{1}$ and $\mathbf{m}_{2}$. Similarly, adding another source $\mathbf{s}_{2}$ only stimulates $\mathbf{m}_{2}$ and it again makes the activities at two measurement sites anticorrelated. This negative correlation is exactly shown by our predicted result in Fig.~\ref{fig:mulit_corr_fn}(g). It displays a negative peak at $\tau=0$.

\subsection{Two dimensional correlations due to a single source}
\begin{figure}[h]
    \centering
    \includegraphics[width=0.5\textwidth]{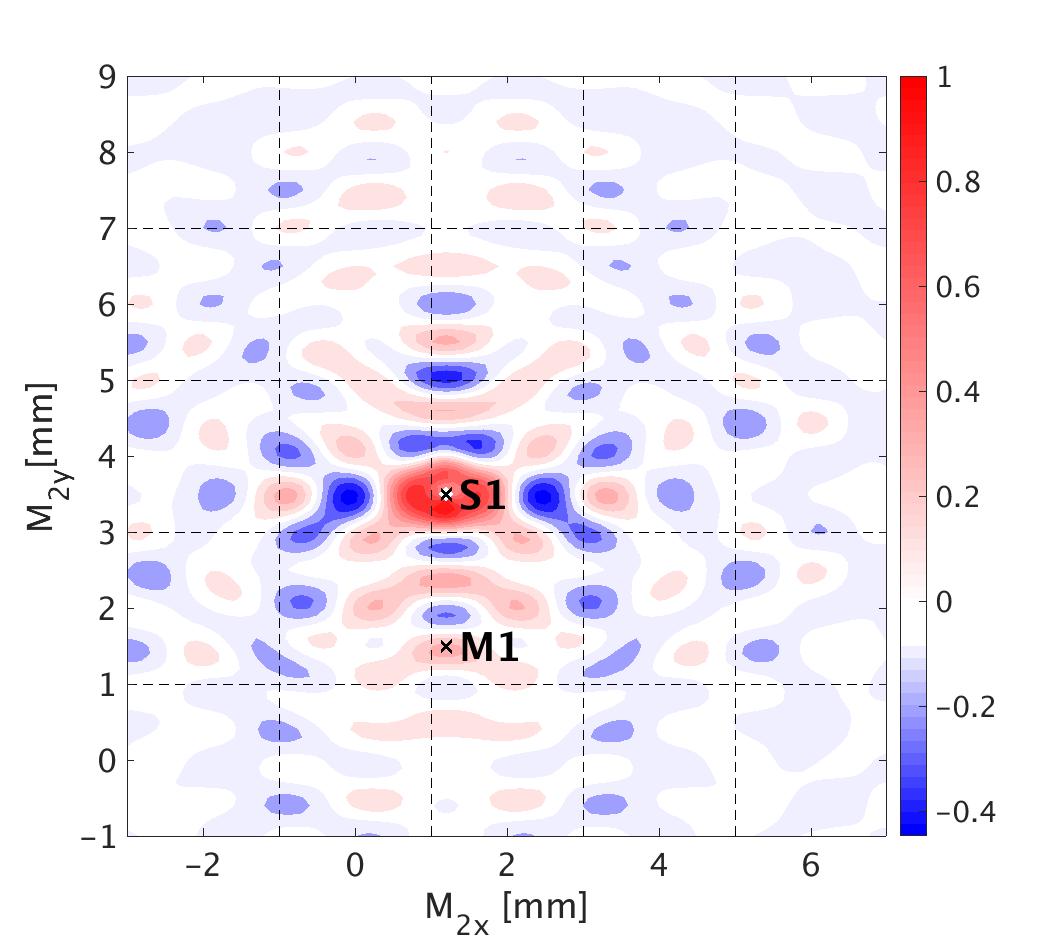}
    \caption{Normalized contour plot of $C(\mathbf{m}_{1}, \mathbf{m}_{2}, 0)$ on V1, from  Eq.~\eqref{eq_corr_m1_m2_numerical} with a single input at $\mathbf{s}_{1}$. The locations of $\mathbf{s}_{1}$, and measurement point $\mathbf{m}_{1}$ are fixed and the location of measurement point $\mathbf{m}_{2}(x,y)$ is given by the axes. The location and OP of $\mathbf{s}_{1}$ and $\mathbf{m}_{1}$ are the same as shown in Figure \ref{fig:mulit_corr_fn}(a). The color bar indicates the strength of the correlation. Dashed lines bound unit cells.}
    \label{fig:corr_plot_vary_m2_one_input}
\end{figure}

To demonstrate how the correlation strength is influenced by the location of the measurement sites and their OP, we fix the location of a source $\mathbf{s}_{1}$ and a measurement point $\mathbf{m}_{1}$, as in Fig.~\ref{fig:mulit_corr_fn}(a). We then map the correlation with the second measurement point ${\bf m}_2$ at $\tau=0$ as a function of the latter's position on V1. The resulting map is shown in Fig.~\ref{fig:corr_plot_vary_m2_one_input}, normalized to the maximum value of $C(\mathbf{m}_{1}, \mathbf{m}_{2}, 0)$.

Figure \ref{fig:corr_plot_vary_m2_one_input} shows that: (i) The strongest positive correlations are located along a vertical axis passing through the source point $\mathbf{s_1}$ whose OP is $90^\circ$;
(ii) Patterns of the correlated regions are almost symmetric around the vertical axis in (i);
(iii) The correlation strength falls off with distance between the two measurement points, as expected from Eq.~\eqref{eq_shape_fn};
In addition, the correlation nearly vanishes when the measurement sites are greater than \SI{7}{mm} apart, and this agrees with the experimental results, which suggested that oscillatory cross-correlations are not observed when the spatial separation of neurons exceeds \SI{7}{mm}.
(iv) The central peak shows that when the distance between $\mathbf{m}_{2}$ and $\mathbf{s}_{1}$ is less than \SI{0.5}{mm}, the correlations are strong and do not depend on the OPs at these locations, in accord with experiments  \citep{bosking_orientation_1997,engel_inter-columnar_1990, gray_oscillatory_response_1989,swindale_review_1996}. (v) The positive correlations correspond to regions of OP approximately equal to $\mathbf{s}_{1}$'s OP. while negative correlation regions correspond to OPs approximately perpendicular to the source OP angle. This shows that only neurons with similar OP to the source respond to the input stimulus.

\subsection{Two dimensional correlations due to two sources}
Here we explore the dependence of the correlation function $C(\mathbf{m}_{1}, \mathbf{m}_{2}, 0)$ on the position of measurement point $\mathbf{m}_{2}$ with two inputs $\mathbf{s}_{1}$ and $\mathbf{s}_{2}$. The location of the measurement points and source points are set up exactly as in the previous case and the additional source $\mathbf{s}_{2}$ has the same OP as $\mathbf{s}_{1}$ (i.e. $90^\circ$).

\begin{figure}
    \centering
    \includegraphics[width=0.7\textwidth]{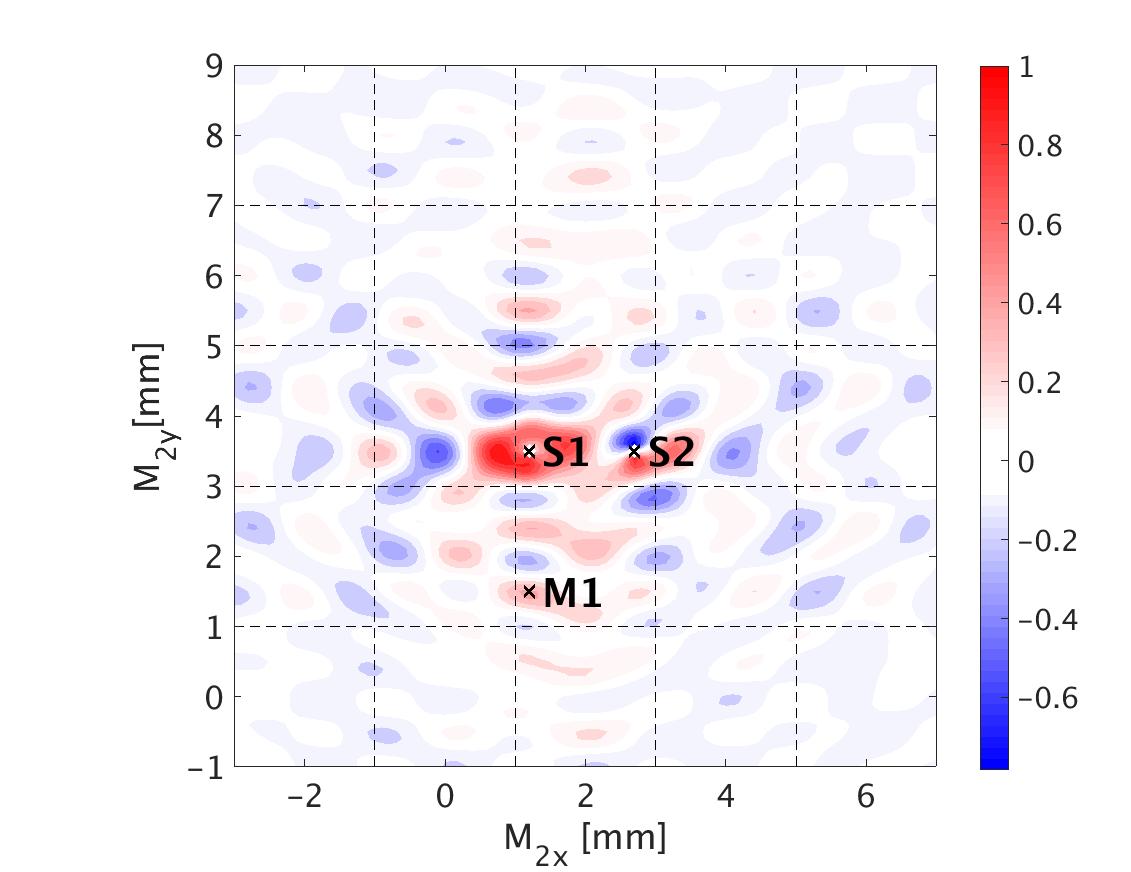}
    \caption{Normalized contour plot of $C(\mathbf{m}_{1}, \mathbf{m}_{2}, 0)$ on V1, from Eq.\eqref{eq_corr_m1_m2_numerical} with two inputs $\mathbf{s}_{1}$ and $\mathbf{s}_{2}$. The locations of the sources and measurement points $\mathbf{m}_{1}$ are fixed and the location of measurement point $\mathbf{m}_{2}(x,y)$ is given by the axes. The location and OP of $\mathbf{s}_{1}$, $\mathbf{s}_{2}$, and $\mathbf{m}_{1}$ are the same as shown in Figure \ref{fig:mulit_corr_fn}(a). The color bar indicates the strength of the correlation function. Dashed lines bound unit cells.}
    \label{fig:corr_plot_vary_m2_two_inputs}
\end{figure}

The resulting map is shown in Fig.~\ref{fig:corr_plot_vary_m2_two_inputs} and has similar properties to the previous case with one input, namely, the strongest correlations between the measurements points are along a vertical axis, which matches the OP of the sources. The positive correlation regions along this axis have a spatial period of \SI{1}{mm}, corresponding to the minimum distance between regions having the same OP angle as the sources. However, the negative correlation regions now tend to align horizontally, which represents the direction orthogonal to the OP. The input source $\mathbf{s}_{2}$ is not surrounded by positive correlation regions as $\mathbf{s}_{1}$ is; rather, the negative correlations right above $\mathbf{s}_{2}$ correspond to a region where the OP of $\mathbf{m}_{2}$ is $\sim 0^\circ$. This is consistent with Sec.~\ref{subsec:temporal_corr}, where we showed that measurement points with orthogonal OPs tend to be anticorrelated at $\tau=0$. In that case, we have predicted that when the OP of two measurement points are $0^\circ$ and $90^\circ$ respectively, the source that is optimal to one of the measurement site introduces negative correlation between the two.

\section{Comparison Between Theory and Experiment}
\label{sec:compare_results}
In this section, we compare the predicted correlation functions  with experimental correlations obtained from \cite{engel_inter-columnar_1990}, who published temporal correlation functions of MUA and LFP data under various conditions.

\subsection{Description of the Experiments}
In these experiments, the MUA and LFP measurements were recorded from an array of electrodes that were inserted in 5 to 7 spatially separated sites in area 17 of anesthetized adult cats, with neighboring recording sites spaced 400 -- 500 $\mu$m apart. Oriented light bars were used as binocular stimulation. Each trial lasted for 10 seconds and one trial set was composed of 10 trials with identical stimuli. During each trial, the light bars were projected onto a screen that was placed \SI{1.10}{\m} in front of the eye-plane of the cat. The autocorrelation function (ACF) and cross-correlation function (CCF) of the MUA data were computed. CCFs were calculated on each individual trials first, then averaged to get the final single CCF corresponding to a specific input stimulus \cite{engel_inter-columnar_1990}.

\subsection{Mapping experimental conditions to a regular lattice}
\label{experimental_cond}
The experimental stimulation was binocular, so a single moving light bar at a specific point in time, maps to two source points on V1 ($\mathbf{s}_{1}$ and $\mathbf{s}_2$), both with OP equal to the bar orientation, one located in left OD column and one in the right OD column.

In Engel {\it et al.}'s experiments, there are five fixed measurement points labeled as $\mathbf{m}_{1}$ to $\mathbf{m}_{5}$. Cells at measurement points $\mathbf{m}_{1}$, $\mathbf{m}_{3}$, and $\mathbf{m}_{5}$ have similar orientation preference and are nearly orthogonal to the OP preference of cells at measurement points $\mathbf{m}_{2}$, $\mathbf{m}_{4}$.
We map these points onto the regular grid used in our model, which results in slight distortion ($<0.5$ mm) of the original cortical surface in order to preserve the measurement-point OPs. The OPs of $\mathbf{m}_{1}$ to $\mathbf{m}_{5}$, computed after mapping onto our regular lattice match to the OPs given by the experiments within $1^\circ$.

\begin{figure}
\centering
\includegraphics[width=0.6\textwidth]{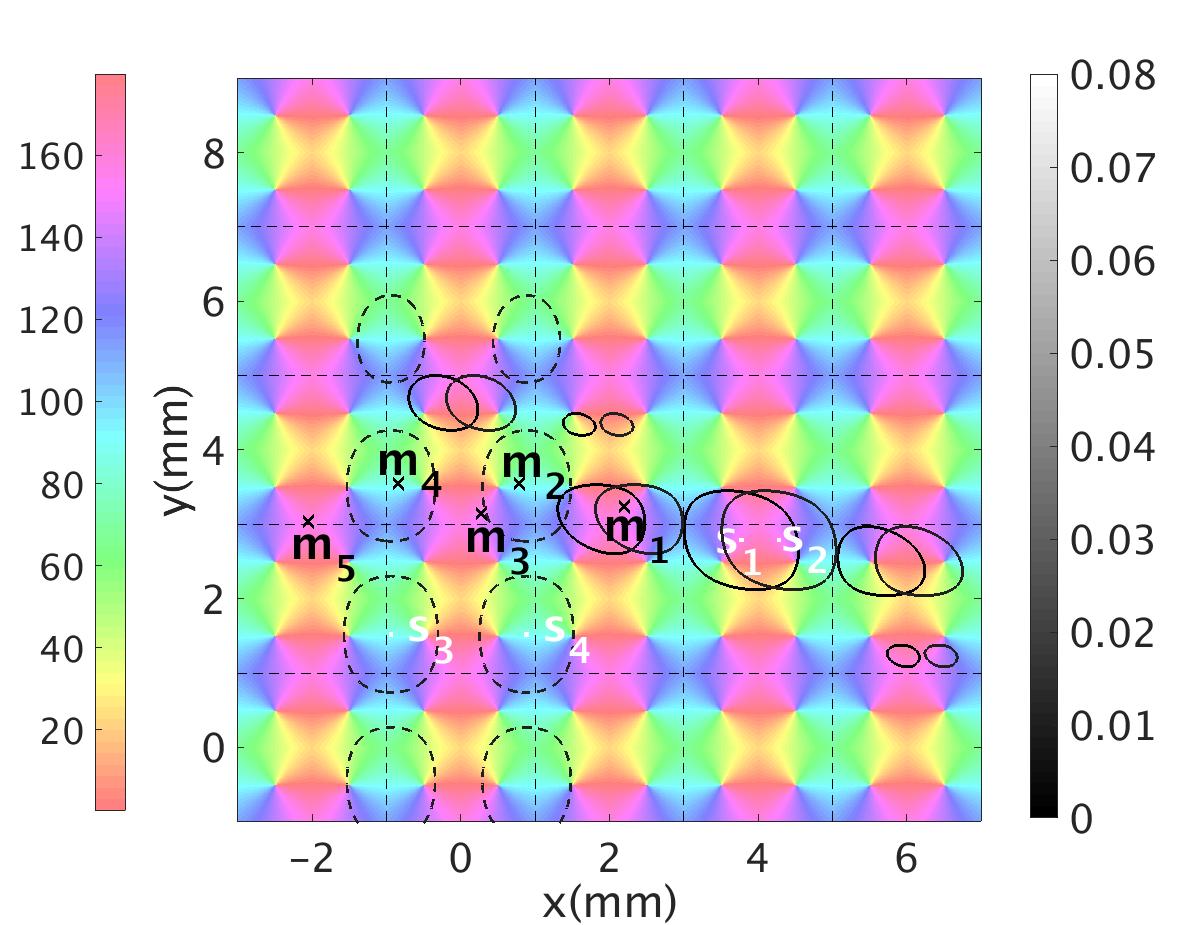}
 \caption{Schematic of the two experimental conditions, showing measurement points $\mathbf{m}_{1}$ to $\mathbf{m}_{5}$ and source points on V1. The first experimental condition corresponds to a stimulus at $157^\circ$ with  corresponding sources denoted $\mathbf{s}_1$ and $\mathbf{s}_2$, with solid gray contours showing the propagation. The second experimental condition corresponds to the a $90^\circ$ stimulus. Here, $\mathbf{s}_3$ and $\mathbf{s}_4$ are the sources and dotted gray contours show the propagation strength according to the grayscale at right. Dotted vertical and horizontal lines bound unit cells and the color bar shows OP in degrees.}
   \label{fig:experi_condn_illu}
\end{figure}

Here, we calculate the temporal correlation functions for two sets of experimental conditions, where the only difference between the two is the OP of the stimulus. One stimulus is oriented at  $157^\circ$ and another one oriented at $90^\circ$. Figure \ref{fig:experi_condn_illu} shows both the stimulation and measurements sites on the idealized OP map. The sources  $\mathbf{s}_1$ and $\mathbf{s}_2$ indicate the $157^\circ$ stimulus, while $\mathbf{s}_3$ and $\mathbf{s}_4$ represent the the $90^\circ$ stimulus. The locations of the measurement sites $\mathbf{m}_{1}$ to $\mathbf{m}_{5}$ are the same for the two sets of experimental conditions.

\subsection{Comparison of Predicted and Experimental Correlation Functions}

\begin{figure}[h!]
    \centering
    \includegraphics[width=0.9\textwidth]{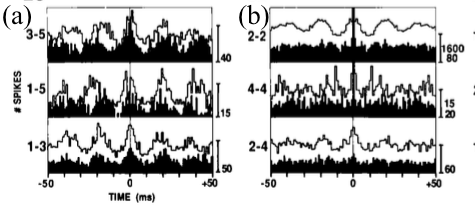}
    \caption{Cross correlograms from experiments recordings calculated by \cite{engel_inter-columnar_1990}. In each case a baseline level of activity shifts the oscillatory part of the correlation upward and must be subtracted for comparison with the theoretical results. (a) Cross correlograms between measurement sites $\mathbf{m}_3$ and $\mathbf{m}_5$, $\mathbf{m}_1$ and $\mathbf{m}_5$, and $\mathbf{m}_1$ and $\mathbf{m}_3$ corresponding to an input light bar oriented at $157.5^\circ$. (b) Auto correlograms of $\mathbf{m}_2$ and $\mathbf{m}_4$ in the top two row, and cross correlograms  on the bottom row between $\mathbf{m}_2$ and $\mathbf{m}_4$, corresponding to the vertical light bar.}
    \label{fig:experiment_result}
\end{figure}

According to the experimental findings in \citet{engel_inter-columnar_1990}, when the input light bar is oriented at  $157.5^{\circ}$, measurement sites $\mathbf{m}_{1}$, $\mathbf{m}_{3}$, and $\mathbf{m}_{5}$ have synchronized oscillatory responses; and, when the input light bar is oriented at $90^{\circ}$, $\mathbf{m}_{2}$ and $\mathbf{m}_{4}$ are stimulated simultaneously. Figure \ref{fig:experiment_result} shows the CCFs and ACFs calculated from the experimental data. In Figure \ref{fig:experiment_result}(a), the synchronized activities at $\mathbf{m}_{1}$, $\mathbf{m}_{3}$, and $\mathbf{m}_{5}$ are evoked by a $157.5^{\circ}$ oriented stimulus. All the cross correlograms are peaked at zero time-lag and have an average oscillation frequency of $\sim$\SI{54}{Hz}. The envelope of the correlograms decreases to $1/e$ of its center peak value at around \SI{45}{ms}. The ACFs and CCF of $\mathbf{m}_{2}$ and $\mathbf{m}_{4}$ from a vertical light bar stimulus are shown in Figure \ref{fig:experiment_result}(b). The CCF between $\mathbf{m}_{2}$ and $\mathbf{m}_{4}$ oscillates at around \SI{55}{Hz}, and it takes more than \SI{50}{ms} for the correlation strength to decrease to $1/e$ of its maximum.

Moreover, in the experiments it was also found that the correlation strength between $\mathbf{m}_{1}-\mathbf{m}_{5}$ is weaker than that between $\mathbf{m}_{1}-\mathbf{m}_{3}$ and between $\mathbf{m}_{3}-\mathbf{m}_{5}$ (i.e., the bar that indicates the number of spikes, on the right of the plot in the second row of Figure~ \ref{fig:experiment_result}(a), has smaller number than other two plots). This is due to the fact that the spatial distance between $\mathbf{m}_{1}$ and $\mathbf{m}_{5}$ is the largest, and the correlation strength falls off with distance.

\begin{figure}[h!]
\centering
\includegraphics[width=0.4\textwidth]{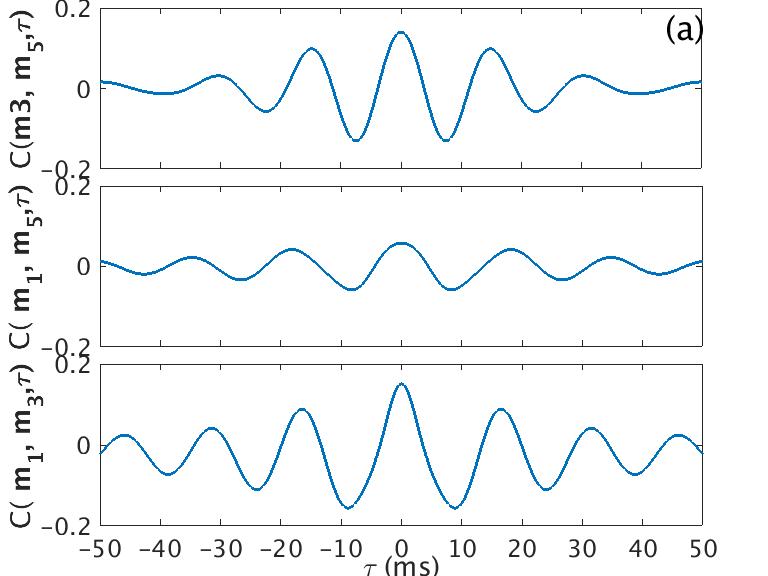}
\includegraphics[width=0.4\textwidth]{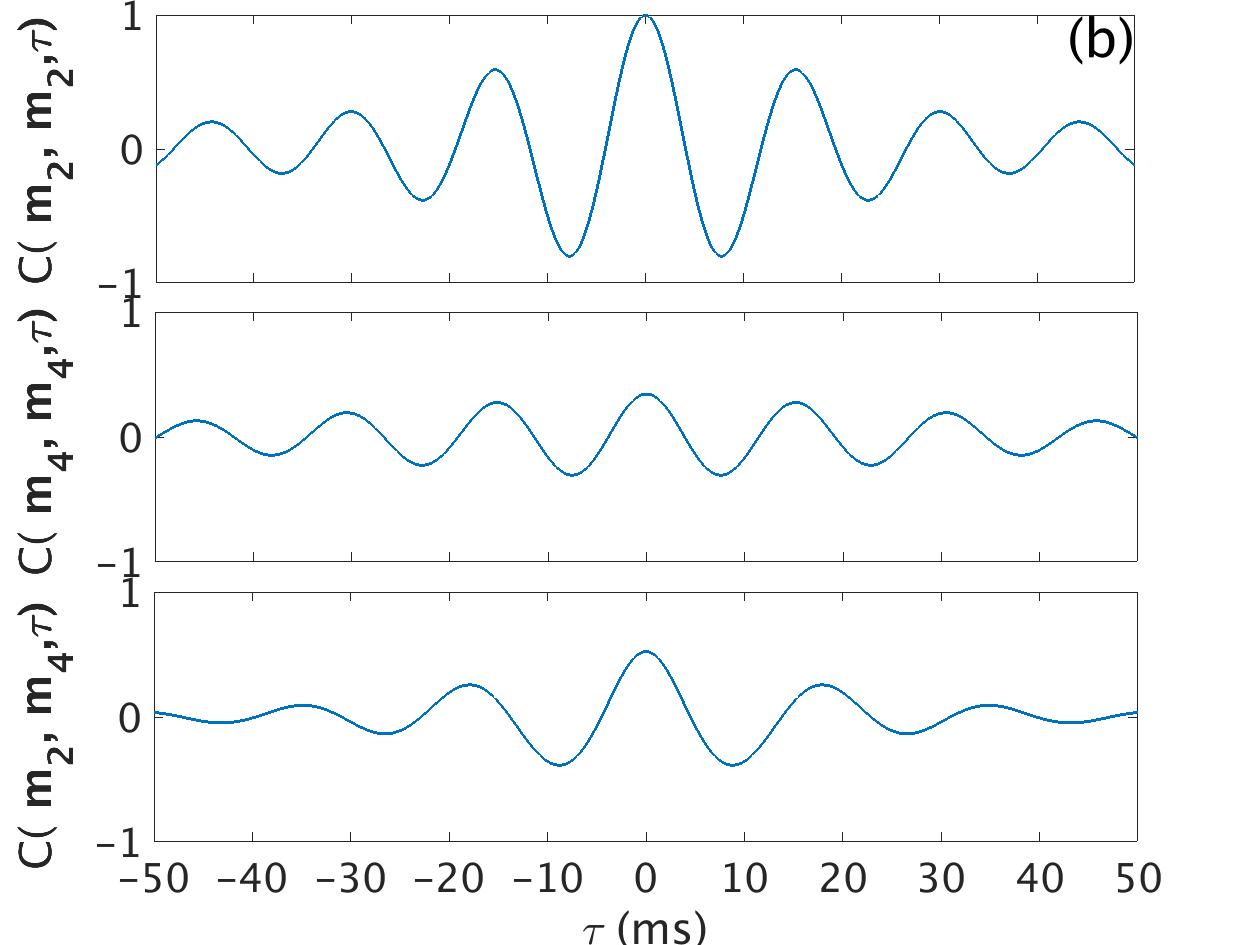}
 \caption{Normalized temporal cross correlation with zero mean for experimental conditions with stimuli at $157.5^\circ$ and $90^\circ$. (a) Normalized temporal correlation between measurement sites $\mathbf{m}_3$ and $\mathbf{m}_5$, $\mathbf{m}_1$ and $\mathbf{m}_5$, and $\mathbf{m}_1$ and $\mathbf{m}_3$ for a stimulus at $157.5^\circ$. Such condition is illustrated in Figure~\ref{fig:experi_condn_illu}. (b) Normalized temporal cross correlation between $\mathbf{m}_2$ and $\mathbf{m}_4$, and the autocorrelation function at $\mathbf{m}_2$ and $\mathbf{m}_4$, for a stimulus at $90^\circ$.}
   \label{fig:predicted_result_from_experiment_1}
\end{figure}

We next explore the properties of our predicted correlation functions using Eq.~\eqref{eq_corr_m1_m2_numerical} with the experimental conditions. Figure \ref{fig:predicted_result_from_experiment_1}(a) shows the plots of our predicted temporal correlation functions between $\mathbf{m}_{3}$ and $\mathbf{m}_{5}$, $\mathbf{m}_{1}$ and $\mathbf{m}_{5}$, and $\mathbf{m}_{1}$ and $\mathbf{m}_{3}$. Similarly to the experimental CCFs, all the theoretical CCFs: (i) are oscillatory and peak at zero time lag;
(ii) have an oscillation frequency around \SI{57}{Hz}; and, (iii) have their characteristic time for the correlation envelope to decrease by $1/e$ of the maximum value at approximately \SI{40}{ms}.
These theoretical results agree with the experimental results, once a nonzero mean baseline is subtracted from the latter.

Our prediction also captures
the spatial dependence of the maximum correlation strength. The plot in the middle row of Figure~\ref{fig:predicted_result_from_experiment_1}(a) corresponds to the correlation between $\mathbf{m}_{1}-\mathbf{m}_{5}$ and has the smallest amplitude among the three CCFs.

Figure \ref{fig:predicted_result_from_experiment_1}(b) shows the predicted temporal correlation function generated by the vertical input light bar. In order to be consistent with the experimental results shown in Figure \ref{fig:experiment_result}(b), the autocorrelation functions of $\mathbf{m}_2$ and $\mathbf{m}_4$ are also included in the top two rows of Figure \ref{fig:predicted_result_from_experiment_1}(b). Both ACFs show oscillations in the gamma band.
The CCF between $\mathbf{m}_2$ and $\mathbf{m}_4$ shows a center peak at $\tau=0$ and oscillates at \SI{55}{Hz}. The time for the envelope decay to $1/e$ of the center peak value is \SI{25}{ms}. These properties are also in line with the experimental findings.

\pagebreak

\section{Summary and Conclusion}
\label{sec:discussion}

We have generalized the spatiotemporal correlation functions in two dimensions that incorporates the spatial structure of the OP map and OD columns of V1. Our results show that the neural activities are synchronized in gamma band when neurons have similar feature preference. The main results are:

(i) The derivation of a shape function that modulates the spatial patchy propagation of the neural signals. The shape function models the propagation such that the orientation of the propagation direction is aligned with the OP of source, and the connected neurons are patchy and periodically located. The parameters of the shape function are tuned to match the propagation ranges observed in experiments \citep{bosking_orientation_1997}.

(ii) The systematic characterization of the 2D two-point temporal correlation function. The generalized correlation function is evaluated numerically for various
combinations of stimulation and measurement sites. The results demonstrate a synchronized gamma oscillation exists between two groups of neurons that have similar OP to the sources. The correlation strength is larger for inter-columnar connections than for intra-columnar connections. As the measurement points are further away from the sources, the correlation strength decreases, and is negligible when the spatial separation of the measurement points exceeds \SI{7}{mm}.

(iii) The construction of a 2D correlation maps. These maps show the changes expected in the peak correlation strength with respect to the variation of the OP of one of the measurement sites, and its distance to a second measurement site. The positive correlations appear as patches on an axis oriented at the OP of the source; and, negative correlations occur where the OPs of the measurement sites are orthogonal to the OP of the source.

(iv) The comparison of the predicted temporal correlations using experimental conditions. Our theoretical results are compared with the experimental findings and shows there is a close match between both in terms of the oscillation frequency and the characteristic decay time of the correlation function envelope. In addition, our CCFs also capture the spatial dependence of correlation strength, which decreases with distance between the measurement sites.

Overall, our generalized spatiotemporal correlation function reproduces the gamma band oscillations observed in V1 and relates the spatially distributed neural responses to the periodic spatial structure of OP and OD in V1. This study lays the foundation to further investigate other visual perception phenomena such as the binding problem.

Future work will focus on using a more realistic lattice of pinwheels and introduce asymmetries between the left/right OD columns to account for strabismus.

\section*{Acknowledgements}
This work was supported by the Australian Research
Council under Laureate Fellowship grant FL1401000025,
Center of Excellence grant CE140100007, and Discovery Project grant DP170101778.

\bibliography{shorttitle,mybib}

\end{document}